\documentclass[aps,pr,reprint,superscriptaddress]{revtex4-1}
\usepackage{dcolumn}
\usepackage{bm}
\usepackage{xcolor}
\usepackage{multirow}
\usepackage{graphics}\usepackage{epsfig}
\usepackage{footnote}
\usepackage{tabularx}
\usepackage[bookmarks=false]{hyperref}
\usepackage[utf8]{inputenc}
\usepackage[english]{babel}
\usepackage{amssymb}
\usepackage{amsmath}
\usepackage{graphicx}
\usepackage{color}
\usepackage{physics}
\usepackage{float}
\renewcommand{\vec}[1]{\boldsymbol{#1}}
\graphicspath{{Figs/}}
\newcommand{\be}{\begin{equation}}
\newcommand{\ee}{\end{equation}}
\newcommand{\bea}{\begin{eqnarray}}
\newcommand{\eea}{\end{eqnarray}}

\def\nn{\nonumber}

\def\pref{\eqref}

\begin{document}

\title{
The electronic structure of a doped Mott-Hubbard surface}

\author{Mattia Iannetti}
\email{mattiaiannetti@gmail.com}
\affiliation{Dipartimento di Scienze Fisiche e Chimiche, Universit\'a degli Studi dell’Aquila, Via Vetoio 10, I-67100 L’Aquila, Italy}
\author{Silvio Modesti}
\affiliation{Dipartimento di Fisica, Universit\'a degli Studi di Trieste, Via Valerio 2, I-34127 Trieste, Italy}
\author{Giovanni Di Santo}
\affiliation{Elettra Sincrotrone Trieste, Strada Statale 14 km 163.5, 34149 Trieste, Italy}
\affiliation{Consorzio Interuniversitario Nazionale per la Scienza e Tecnologia dei Materiali, Unità di ricerca di Trieste, Strada Statale 14 km 163.5, 34149 Trieste, Italy}
\author{Marco Caputo}
\affiliation{Elettra Sincrotrone Trieste, Strada Statale 14 km 163.5, 34149 Trieste, Italy}
\author{Polina M. Sheverdyaeva}
\affiliation{Consiglio Nazionale delle Ricerche, Istituto di Struttura della Materia (CNR-ISM), SS 14, km 163,5 I-34149 Trieste, Italy}
\author{Paolo Moras}
\affiliation{Consiglio Nazionale delle Ricerche, Istituto di Struttura della Materia (CNR-ISM), SS 14, km 163,5 I-34149 Trieste, Italy}
\author{Fabio Chiapolino}
\affiliation{Physics Department, Technical University of Munich, James-Frank-Strasse 1, 85748 Garching bei München, Germany}
\author{Tommaso Cea}
\affiliation{Dipartimento di Scienze Fisiche e Chimiche, Universit\'a degli Studi dell’Aquila, Via Vetoio 10, I-67100 L’Aquila, Italy}
\author{Cesare Tresca}
\affiliation{CNR-SPIN C/o Dipartimento di Scienze Fisiche e Chimiche, Universit\'a Degli Studi dell'Aquila, Via Vetoio 10, I-67100 L'Aquila,  Italy}
\author{Erio Tosatti}
\affiliation{International School for Advanced Studies (SISSA), 34136 Trieste, Italy}
\affiliation{The Abdus Salam International Centre for Theoretical Physics (ICTP), 34151 Trieste, Italy}
\author{Gianni Profeta}
\affiliation{Dipartimento di Scienze Fisiche e Chimiche, Universit\'a degli Studi dell’Aquila, Via Vetoio 10, I-67100 L’Aquila, Italy}
\affiliation{CNR-SPIN C/o Dipartimento di Scienze Fisiche e Chimiche, Universit\'a Degli Studi dell'Aquila, Via Vetoio 10, I-67100 L'Aquila,  Italy}


\begin{abstract}
The  Sn/Si$(111)-(\sqrt{3}\times\sqrt{3})$R$30^\circ$  surface, a 2D Mott insulator, has long been predicted and then found experimetally to metallize and even turn superconducting upon boron doping. In order to clarify the structural, spectroscopic and theoretical details of that evolution, here we present ARPES data supplementing  morphology and scanning tunneling measurements. These combined experimental results are compared with  predictions from a variety of electronic structure approaches, mostly  density functional DFT+U, but not neglecting Mott-Hubbard  models, both ordered and disordered. These theoretical pictures address different spectroscopic aspects, including the 2D Fermi surface, the Hubbard bands, etc. While no single picture account for all observations at once, the emergent hypothesis compatible with all data is that metallization arises from sub-subsurface boron doping, additional to the main standard subsurface boron geometry, that would leave the surface insulating. These results advance the indispensable frame for the further understanding  of this fascinating system.
\end{abstract}

\maketitle

\section{Introduction}

\begin{figure*}
\centering
\includegraphics[width=2\columnwidth]{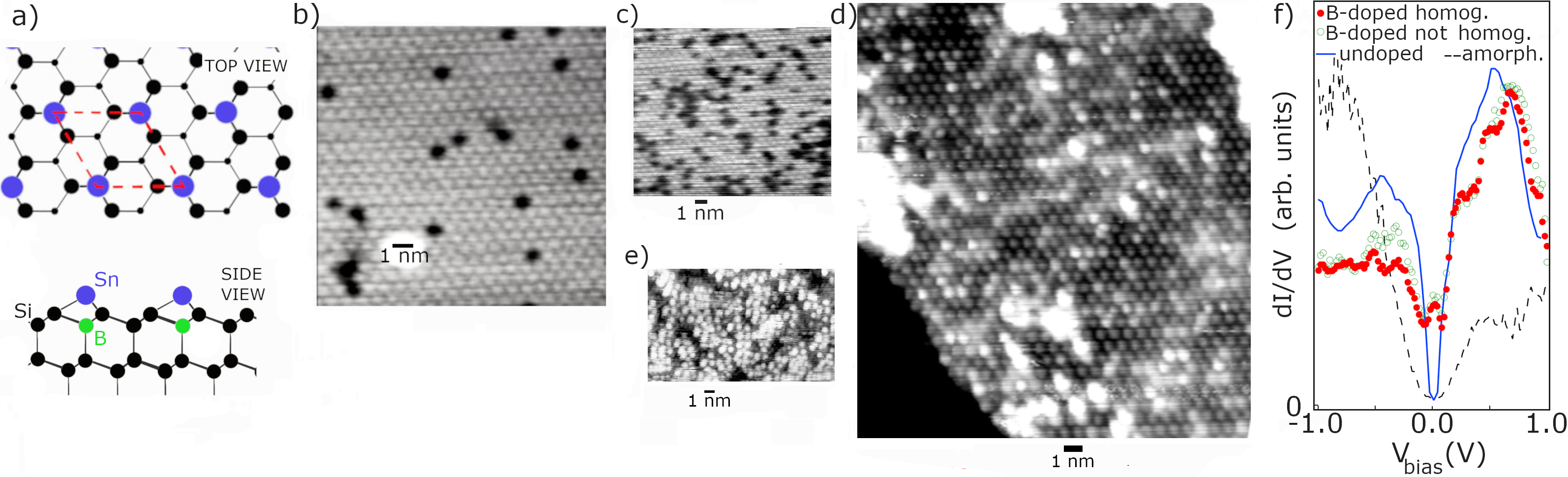}
\caption{
a) Sketch of the B-doped Sn/Si$(111)\text{-}(\sqrt{3}\times\sqrt{3})\mathrm{R}30^\circ$ structure. At saturation ($0.33$ ML of B), all the sites below the Sn atoms (S$_5$ sites) are occupied by substitutional B atoms. At lower doping, both Si and B occupy the same S$_5$ sites randomly. Si substitutes all the B atoms in the undoped samples. The unit cell is marked in red ;
b) STM topographic images of homogeneous Sn-$\sqrt{3}$ regions with low defect density (likely substitutional Si adatoms, dark spots in the image);
c) regions with high defect density;
d) regions with inhomogeneous doping, the white areas are amorphous clusters;
e) amorphous regions (see text);
f) tunneling spectra of the well-ordered homogeneous and the inhomogeneous (disordered) p-doped surface, of the pristine B-free, Sn-$\sqrt{3}$ surface on an n-doped substrate, and of the amorphous regions at 65 K. The set points are $I = 30$ pA and $V_\text{bias} = 0.6$ V for panels b) to f).
}
\label{fig1new}
\end{figure*}

Phenomena typical of two-dimensional  (2D) or nearly two-dimensional correlated electron systems, such as unconventional superconductivity \cite{Wu_2020,Biderang_2022,Wolf_2022,Cao_2018, nakamura_2018, Machida_2022}, pseudogap \cite{Xiong_2022}, metal-insulator transitions \cite{Modesti_2007,Cortes_2006,Profeta_2007,Odobescu_2017}, commensurate charge density waves \cite{Tosatti_1976, Fazekas_1979, Tresca_2018,Cortes_2013} and unusual magnetic ordering \cite{Li_2013,Lee_2013}, have been observed in triangular adatom lattice on semiconductor surfaces.\\
From the structural and electronic point of view these 2D systems are  much simpler than other 3D strongly correlated materials such as, e.g., cuprates.
For the latter there is in fact less general consensus on the phases generated upon doping the parent Mott-Hubbard insulator.
Triangular adatom layers on semiconductors represent therefore good  models where a better understanding of the interplay between Mott-Hubbard state, superconductivity, charge ordering, magnetism, geometric frustration, and doping can hopefully  be attempted. \\ 
One third of a monolayer (ML) of Sn on the Si$(111)$ surface in the Sn/Si$(111)-(\sqrt{3}\times\sqrt{3}) $R$30^\circ$ structure (Sn-$\sqrt{3}$ in the following) is a good  example of nominally sigle-band correlated electron system on a 2D triangular lattice. Each Sn adatom occupies a T$_4$ site on the Si$(111)$ surface \cite{Conway1989} (see Fig.\ref{fig1new}a).
Three of the adatom's valence electrons  bind to three underlying Si surface atoms
The remaining unpaired electron is donated to a surface state band in the Si energy gap, a band that becomes exacly half-filled in the  $\sqrt{3}\times\sqrt{3}$ adatom coverage. 
Owing to the large Sn-Sn distance ($6.7 \AA$) the band is a mere $\sim 0.3$ eV wide.
The on-site Coulomb repulsion of at least $1$ eV \cite{Profeta_2007,Wu_2020} therefore makes this state strongly correlated and prone to electronic and structural instabilities. The physical properties of Sn/Si were first investigated looking for a possible CDW or charge disproportionated  ground state\cite{Ballabio_2002} -- none of which occurs in this case --  and then for electronic instabilities such as a Mott-Hubbard transition \cite{Profeta_2007,Hansmann_2013}. 
Very similar systems like Sn/Ge(111) \cite{Sanna_2019,Schuwalow_2010,Ortega_2002,Lee_2013,Profeta_2007,Cortes_2006,Cortes_2013,LeLay_applsurfsc98}, Pb/Si(111)\cite{Tresca_2018,Cudazzo_2008,Ren_prb16, Brihuega_prl2005} and Pb/Ge(111)\cite{Goldoni_prb97,Mascaraque_prb98,LeLay_applsurfsc98,Ren_prb16,Tresca2021} all display a $\sqrt{3}\times \sqrt{3}$ reconstruction at ambient temperature, with a large variety of low-temperature ordered phases ranging from CDW distortions and disproportionations to spin-chiral bands and metal-insulator transition\cite{Santoro_prb99}.
Although vastly studied since 1996\cite{Carpinelli_nature96}, the experimental and theoretical investigations continued because of this richness. 
Most exciting, a superconducting phase in doped Sn-$\sqrt{3}$, first predicted in 2007 to arise from a Mott insulating phase \cite{Profeta_2007} was recently discovered experimentally \cite{Ming_2017}.\\
In past experiments, undoped Sn-$\sqrt{3}$ was confirmed to have  a Mott-Hubbard insulating ground-state when undoped, and to undergo an insulator-metal transition  with increasing temperature \cite{Modesti_2007,Profeta_2007,Odobescu_2017}.  
As anticipated theoretically \cite{Profeta_2007}, hole-doping  drives the insulator back to a  strange metal phase \cite{Ming_2017}  characterised by a pseudogap  \cite{Xiong_2022}.  This phase, obtained by boron doping, eventually becomes superconducting below about $4$ K\cite{Wu_2020}. The main site occupied by B acceptors is the subsurface site of the Si substrate (see Fig.\ref{fig1new}a) \cite{Ming_2017}, capturing the Sn-p$_z$ dangling bond electron from the surface band \cite{Profeta_2007}.
Theoretical studies (inclusive of long-range Coulomb  electron-electron interactions in an extended Hubbard model whose  parameters were obtained by ab initio calculations)  predict an unconventional superconducting (SC) phase which could change from chiral $p$-wave to chiral $d$-wave to $f$-wave as a function of doping and strength of the non-local interactions \cite{Biderang_2022,Wolf_2022,Iannetti_2023}.
The detailed understanding of the nature of the metallic B-doped phase, which hosts the superconducting phase, is still limited and deserves further investigation. 
Important questions concerning the electronic properties of the doped system are still without  a definitive answer. 
Experimental information about the B-doped metal was obtained so far only with local probes like Scanning Tunneling Microscopy (STM) and Spectroscopy (STS). These tools reveal a coexistence of distinct microscopic phases. 
The ordered and homogeneous phase from which superconductivity arises is expected to be characterized by an almost rigid shift of the Fermi energy with doping  \cite{Wu_2020,Ming_2017}. This scenario is usually employed to describe the hole-doping regime of Mott-insulators\cite{Cao_2018,Ming_2017,Wolf_2022,Kim_2023}, thus disregarding the possible chemical role of dopants.

 
 In order to understand if and how the rigid-band model properly  describes the B-doped Sn/Si, it is crucial to extend the investigation of the  electronic structure of the doped phase by nonlocal probes such as Angle-Resolved Photoemission Spectroscopy (ARPES) and compare them to first-principles, translationally invariant  theoretical calculations.
That is the scope of this work, where we address the electronic properties of  B-doped  Sn-$\sqrt{3}$ using ARPES supported by local probes of the surface like STS/STM, and through    realistic first-principles calculations of the doped phases and model calculations where strong electron correlations and doping inhomogeneities will be treated.

\section{Experimental results and discussion}
\label{Experimental}



The samples  
were prepared by flashing B-doped  Si(111) wafers ($\rho =0.002$ $\Omega$cm and $1$ $\Omega$cm) above $1200$ $^\circ$C by electron bombardment for the photoemission measurements and by direct current heating for the STM-STS measurements. The B density in the subsurface region was controlled by annealing the samples at $\simeq 950$ $^\circ$C  for different duration and was measured by STM and by B1s core level photoemission with a relative error of about $30$\%. The B coverage of 0.33 ML corresponds to saturation doping, $i. e.$ one B atom in the S$_5$ site every 3 Si atoms of the ideal $(1\times 1)$ surface. While most of the B atoms are in the first bilayer from the surface, other B atoms are known to be 
present in deeper layers with lower densities \cite{shen_94, Hirayama_97, chimici}. Sn was dosed on the surface by evaporation from a crucible with the substrate kept at $520$ $^\circ$C. The 0.33 ML Sn coverage was assumed to be that above which the LEED spots of the high-coverage $2\sqrt{3}\times 2\sqrt{3}$ phase start to appear. The STS data were measured at $60$ and $300$ K with a homemade STM-STS system, while the photoemission spectra were acquired at about $30$ K at the VUV photoemission beamline of Elettra. The energy resolution of the photoemission spectra was $\approx 0.08$ eV, mainly caused by the inhomogeneous surface photovoltage broadening. The angle between the photon beam and the surface normal was between $45^\circ$ and $58^\circ$ with the electric field of the radiation in the plane containing the beam and the surface normal. The binding energies are referred to the Fermi level. The surface photovoltage (SPV) effect shifts the spectra by about $0.12\pm 0.02$ eV to lower energies at $30$ K. This shift was evaluated by comparing spectra of the entire Si valence band at 300 and 30 K and was subtracted to the energies of the 30 K spectra to correct the systematic error caused by SPV.

\subsection{Morphology}

At intermediate and high B doping $\Theta_B$ ($0.06~$ ML$<\Theta_B <0.3$ ML) and at Sn coverage of $0.3~$ ML, STM shows the coexistence of  homogeneous Sn-$\sqrt{3}$ regions with low defect density (Fig.\ref{fig1new}b), with less homogeneous Sn-$\sqrt{3}$ regions with inequivalent Sn adatoms or Sn vacancies (Fig.\ref{fig1new}c and d), and with amorphous regions (Fig.\ref{fig1new}e). These diverse phases were already reported to coexist, and described in Supplementary Materials of Ref. \cite{weit_2018}. The homogeneous Sn-$\sqrt{3}$ phase has Sn adatoms which differs in apparent height by less than 0.02 nm both at positive and negative sample bias and some substitutional Si adatoms and Sn-vacancies (about $4$\% of the surface layer) (Fig.\ref{fig1new}b). The same defects were also found with higher densities in the defected Sn-$\sqrt{3}$ phase (Fig.\ref{fig1new}c). In addition, we find Sn-$\sqrt{3}$ regions showing Sn adatoms with different apparent height (“bright” Sn adatoms, up to about 0.1 nm higher than others for both positive and negative sample biases (Fig.\ref{fig1new}d). These  inhomogeneities  are similar to those observed in the isostructural and isoelectronic Si(111)-B $\sqrt{3}\times\sqrt{3}\text{R}30^{\circ}$ ( Si-$\sqrt{3}$ ) and attributed to an inhomogeneous distribution of subsurface B dopants \cite{Hirayama_97,chimici} (see also Appendix \ref{APPENDIX_EXP} for more details).
In particular, the brighter Sn adatoms, clearly visible in Fig.\ref{fig1new}d, tend to align in 3-7 adatoms long chains without an apparent long range order, a peculiar effect observed also in B-doped Si-$\sqrt{3}$ \cite{lyo_1989,shen_94, Hirayama_97}.
In line with the Si-$\sqrt{3}$ case \cite{lyo_1989,shen_94, Hirayama_97}, we attribute the “bright” protrusions to surface Sn atoms without B dopants below them, thus retaining the p$_z$ electron at the Fermi level making them brighter than the other Sn atoms in the STM images. The vertical displacements of Sn adatoms without B-S$_5$ dopant also contribute to the contrast (see below). This observation indicates that the doping could be local and without long range order. Conversely, the absence of adatom contrast in the homogeneous phase of Fig.\ref{fig1new}b could in principle be explained by a complete B layer below the Sn adatoms, in agreement with what is observed on the saturated Si-$\sqrt{3}$ surface. However, this phase forms even when the average B coverage is well below 0.33 ML. Moreover, we never observed the most common defect always present on a B-saturated surface, i.e. isolated bright adatoms on top of rare vacancies of the B layer. The absence of this inevitable defects, together with spectroscopic evidence presented in the next sections, will induce us to propose a different model for the homogeneous phase.
In passing, we note that the presence of  amorphous regions formed by adatoms spaced by 0.5 - 0.8 nm (Fig.\ref{fig1new}e) is in line with what is observed in Ref.\cite{Ming_2017} (see Appendix \ref{APPENDIX_EXP}) and with similar regions observed on the B-doped Si-$\sqrt{3}$ surface at low doping \cite{Hirayama_97}. 

It should be noted that when 1/3 of a ML of Sn is dosed on the Si surface and annealed at 520 $^\circ$C the Si adatoms that Sn replaces are expected to form about 1/6 of a bilayer on top of the doped Si surface, presumably at the step edges. The Sn adatoms that cover this new regions of the Si surface have B atoms in the second bilayer below them, contrarily to the rest of the Sn adatoms. Since B diffuses appreciably only above 850 - 900 $^\circ$C, little dopant inter-diffusion is expected at the much lower growth temperature of the Sn-$\sqrt{3}$ overlayer, where the mean displacement should be less than 10$^{-2}$ nm according to the bulk diffusion coefficient \cite{Vick, Christensen}. Indeed, experimental data indicates that a Si-$\sqrt{3}$  overlayer can grow on top of the B-doped Si-$\sqrt{3}$ surface without substantial B displacement, i.e. with little B surface segregation, if the temperature is similar to that we used for the sample preparation \cite{Zotov_1996, Headrick_1990,Korobtsov,Schulze}, leaving the dopants in the second bilayer. Therefore, it is likely that one of the phases discussed above forms on regions covered by displaced Si adatoms with most of the B dopants separated from the Sn atoms by an entire Si bilayer. As discussed in the theoretical section and in Appendix B there are evidences that support the presence of this extra Si bilayer. 
Moreover, additional B atoms are present in deeper layers, since there are experimental indications that the concentration of B in the first nm below the first bilayer is high, of the order of $5\times 10^{19}$ cm$^{-3}$ \cite{Makoudi_2017} or above \cite{shen_94}, in agreement with our STM results (see Appendix \ref{APPENDIX_EXP}).

\subsection{Tunneling spectroscopy}
The tunneling spectra reported in Fig.\ref{fig1new}f of homogeneous (red circles) and inhomogeneous hole-doped (green circles) Sn-$\sqrt{3}$ phases (relative to the STM images reported in Fig.\ref{fig1new}b)-d)) at $\sim 65$ K and with $\sim 0.15$ B ML in the subsurface region show two broad peaks in the region $0.3 - 0.7$ eV below and above E$_F$, attributed to the lower and higher Hubbard bands, shifted by no more than $0.1$ eV with respect to the undoped case,  and an additional contribution near the Fermi level, interpreted as a the quasi-particle peak in Ref.\cite{Ming_2017}. We do not emphasize the differences between the two spectra because they are within our experimental error.  
\footnote{The spectra are likely measured with two different nanotips of the same STM tip with presumably slightly different electronic structure and on substrates with B doping that may differ by even 30 \%.} 
We observe that at the same temperature the Sn overlayer on a $n$-doped substrate ($\rho =0.003 $ $\Omega \,$cm), which yields a nearly undoped Sn-$\sqrt{3}$ surface \cite{Ming_2017}, has a 
differential conductivity (blue curve) by at least a factor ten lower at E$_F$ (between the Hubbard bands) \cite{Modesti_2007, Odobescu_2017,Ming_2017},  showing a clear  Mott-Hubbard gap.
The differential conductance at E$_F$ of the competing amorphous phase of Fig.1e measured by STS (dashed line) is less than 1/5 of that of the B-doped Sn-$\sqrt{3}$ phase (Fig.\ref{fig1new}f). This low value suggests that the amorphous phase is insulating and therefore we expect that its contribution  to the photoemission spectra at  E$_F$ should be negligible. Possible small amounts of the insulating $2\sqrt{3}\times2\sqrt{3}$R$30^{\circ}$ Sn phase with filled bands below -0.5 eV \cite{Eriksson} should also contribute very little to the signal at E$_F$, letting the ordered B-doped Sn-$\sqrt{3}$ phase dominate the photoemission spectra around the Fermi level.

\begin{figure}
\centering
\includegraphics[width=\columnwidth]{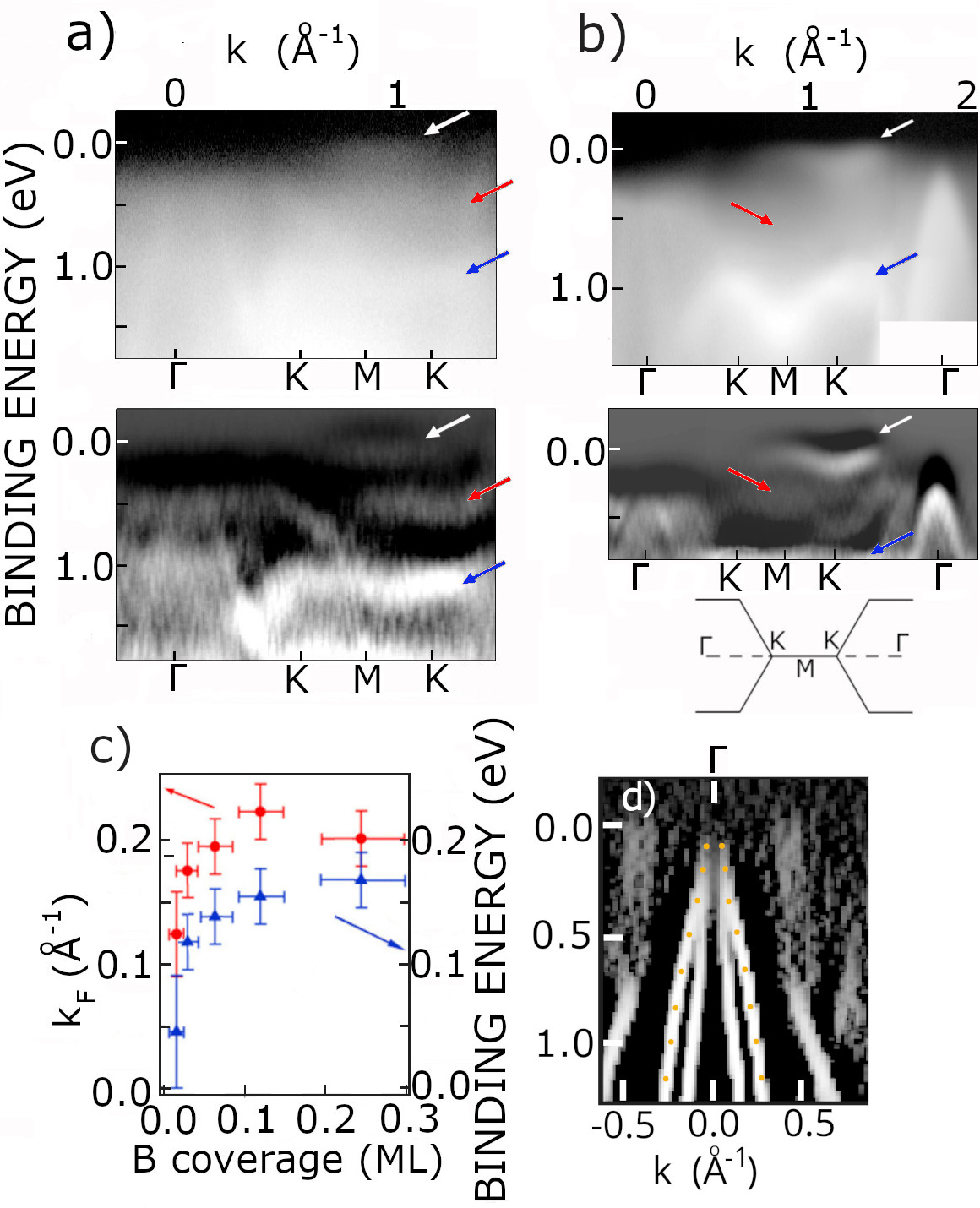}
\caption{
Photoemission intensity along the $\Gamma K$ direction of the Sn-$\sqrt{3}$ BZ for B coverage of about $0.04$ a) and $0.12$ ML b) as a function of binding energy and wavevector k at $30$ K. The panels display the photoemission intensity in logarithmic scale (top) and the opposite of its second derivative with respect to energy (bottom). The inset shows the direction sampled and the Sn-$\sqrt{3}$ BZs. The two $\Gamma$ points of the Sn-$\sqrt{3}$ BZs shown are also $\Gamma$ points of the $1\times 1$ surface BZ. The white, red and blue arrows indicate the B-induced partially filled surface band, the lower Hubbard band, and the backbond bands respectively (see text). c) doping dependence of the Fermi wavevector (measured from K) and of the binding energy of the bottom of the dopant-induced  band. d) top of the Si valence band in the third BZ obtained from the second derivative with respect to k of the photoemission intensity along the $\Gamma$M direction. The plot is in a logaritmic scale. The wavevector is measured from the $\Gamma$ point. The yellow dots are guide to the eye. The band rises at or above the Fermi level (see also Appendix B). 
}
\label{Fig_2}
\end{figure}

\subsection{ARPES}

Fig.\ref{Fig_2} shows the angle resolved photoemission spectra (ARPES) of the B-doped Sn-$\sqrt{3}$ phase at 30 K for $\Theta_B$ 0.04 and 0.12 ML, measured along the $\Gamma K$ direction of the Sn-$\sqrt{3}$ surface unit cell. The features of the intensity maps versus energy and wavevector (top of panels a and b) sit on a strong background caused by the amorphous and disordered phases. This background can be suppressed by taking the second derivative of the intensity with respect to the energy (bottom of panels a and b). Low B doping ($\Theta_B$ $\sim 0.04$ ML) (Fig.\ref{Fig_2}a) does not drastically modify the spectrum of the Mott insulating phase: the dispersive bands within 0.35 \AA$^{-1}$ from $\Gamma$ are the Si valence bands,  brought closer to E$_F$ by the surface doping. 

We observe a weak and broad structure without appreciable dispersion at $0.3-0.4$ eV in some regions of the BZ (see red arrow), particularly evident from the second derivative of the spectra  (lower panel of Fig.\ref{Fig_2}a) 
(see Appendix \ref{APPENDIX_EXP} for raw data analysis).  The energy position of this structure corresponds to the lower Hubbard band observed in the STS data and in the photoemission spectra of the undoped Sn$-\sqrt{3}$ phase \cite{Modesti_2007}. This structure is more evident at these low B doping regimes (see Appendix \ref{APPENDIX_EXP}), and is the very likely the residual of the lower Hubbard band, weakened by the doping. 
A filled surface band disperses from $-1.7$ eV at $\Gamma$ point to $-1$ eV at K point and $-1.3$ eV at M (see Fig.\ref{Fig_2} b, blue arrow). This is similar to the measured and calculated bands of the saturated Si$(111)$-B $\sqrt{3}$ surface which were attributed to Si backbonds and Si-B bonds, and it is also present in the undoped Sn-$\sqrt{3}$ surface originating from the Sn backbonds bands \cite{Grehk_1992, Shi_2002, Aldahhak_2021, PhysRevB.62.1556}. We therefore assign this band to the hybridization between B-Si bonds and the backbonds of the Sn adatoms.

A new and weak additional photoemission intensity appears within $0.1$ eV from the Fermi level near the K and the M points (see the white arrow in Fig.\ref{Fig_2}a). We notice that in the  spectra acquired at higher B doping levels (0.12 ML), reported in Fig.\ref{Fig_2}b, this feature evolves into a shallow electron-like partially filled band centered at the K point of the second Brillouin zone (BZ). Its minimum at K is $0.13 \pm 0.02$ eV below the Fermi level and  crosses E$_F$  $0.18 \pm 0.02$ \AA\ $^{-1}$ from K, as shown in Appendix \ref{APPENDIX_EXP}. \footnote{The errors are caused by broadening of the spectra induced by spatial and temporal dishomogeneity of the surface photovoltage effect, by intrinsic broadening and by disomogeneous B coverage.} 
Interestingly, the energy position of this last band is in the range of the Quasi-Particle Peak (QPP) observed in STS on the well ordered doped phase (see Fig.\ref{fig1new}f and Ref.\cite{Ming_2017, Xiong_2022} ) and  is absent in the undoped system \cite{Modesti_2007, Li_nature2013}. In addition, the band cannot be attributed to neither the insulating $2\sqrt{3}\times 2\sqrt{3}$R30 phase of Sn/Si$(111)$\cite{Eriksson}, which could be a possible minority phase on our surfaces, nor to the amorphous phase as it disperses at the Fermi energy. The effective mass of this band is about $1.0 \pm 0.2$ times electron mass.
We followed the occupancy of this band as a function of the average B coverage, reporting in Fig.\ref{Fig_2}c both its  Fermi wave-vectors ($k_F$) and  energy minimum, estimated from  the photoemission intensity maps at E$_F$ (see below) and by fitting the photoemission spectra at different wavevectors (see Appendix \ref{APPENDIX_EXP}).

For our experimental setup, photon energies ($30-60$ eV) and photon electric field at about $45^\circ$ from the surface in the plane containing the surface normal and photon beam, this partially filled band is evident only at some corners of the second and third BZ. We attribute this to multiple scattering effects on the final-state similarly to what is observed on related surfaces in similar conditions \cite{Modesti_Si,Uhrberg}.

Another difference between the spectra of the B-doped and those of the undoped Sn-$\sqrt{3}$ surface is the energy position of the top of the Si valence band. While it is at least 0.2 eV below E$_F$ without B \cite{Modesti_2007,Lobo_2003}, it reaches or rises above the Fermi level in the doped case. This can be seen in panel d) of Fig.\ref{Fig_2} which reports the intensity distribution in the third BZ along the $\Gamma$M direction.  This rise of the valence band at $\Gamma$ occurs only when the Sn layer forms on the doped surface and is not present on the pristine Si terminated Si(111)-B. In our geometry and photon energy range the photoemission intensity of the top of the valence band is always very weak and is detectable only at high parallel wavevectors (see also Appendix B). 

\begin{figure}
\centering
\includegraphics[width=\columnwidth]{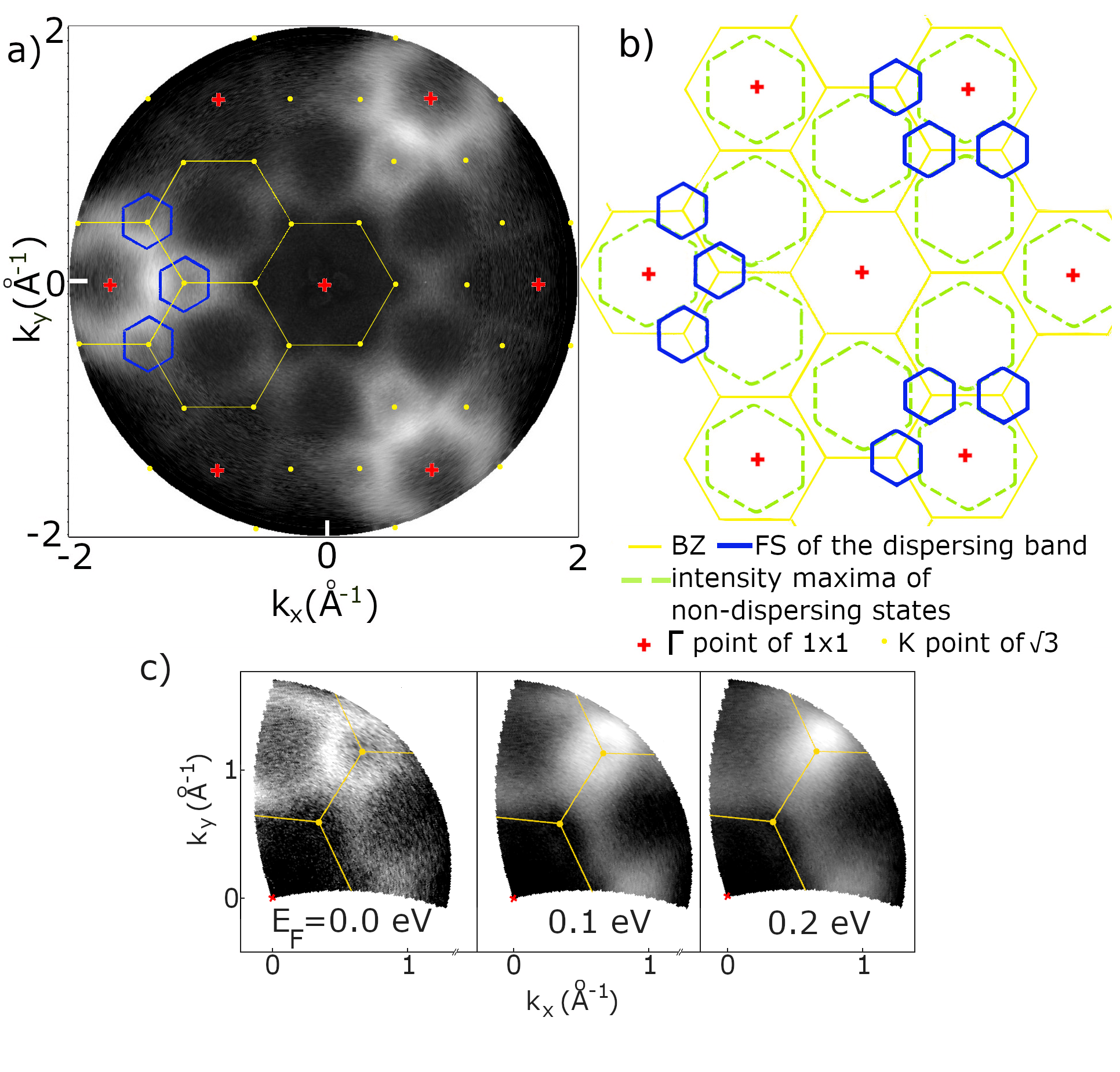}
\caption{
a) Map of the photoemission intensity of Sn-$\sqrt{3}$ at the Fermi level at $30$ K for a B coverage of $0.1$ ML as a function of the parallel wavevector $k$ for $47.5$ eV photon energy. The red crosses indicate the $\Gamma$ points of the ideal 1x1 surface, the yellow points mark the K points of the Sn-$\sqrt{3}$ BZs, the small blue hexagons highlight the contribution from the metallic surface band in part of the map. Bright indicate high intensity, dark low intensity. b) Sketch of the observed maxima of the spectral intensity plotted with the Sn-$\sqrt{3}$ BZs marked in yellow. The small blue hexagons correspond to the Fermi surface of the shallow and partially filled surface band of Fig.\ref{Fig_2}, the large green hexagons to the maxima of the photoemission intensity of the tail of the non-dispersing band at $-0.25$ eV. The size of the latter hexagons depends on the photon energy (see text). c) Map of the photoemission intensity at different binding energies. The BZs are marked in yellow, the yellow points are the K points.
}
\label{Fig_3}
\end{figure}

The map of the photoemission spectral intensity distribution at the Fermi energy at $30$ K is shown in Fig.\ref{Fig_3} as a function of the parallel momentum for a B coverage of  $0.1$ ML and a photon energy of $47.5$ eV. The data acquired in a $150^\circ$ sector have been symmetrized by rotations of $120^\circ$. The map shows contours with small (large) approximately hexagonal shapes centered at $K$ ($\Gamma$) points of the Sn-$\sqrt{3}$ BZs. 
The apothems of the hexagons are $0.22$ \AA$^{-1}$ and  $0.45$ \AA$^{-1}$ for the small and large structures, respectively. 
The smaller hexagonal features mark the Fermi level crossing of the partially filled surface band discussed above. On the contrary, we cannot attribute the larger ones to any dispersing bands observed in ARPES. 
Indeed, maps of the photoemission intensity at different binding energies (ranging from $0$ to $0.5$ eV) show that the dimension of these hexagons does not change within $0.1$ \AA$^{-1}$, while they disappear at higher binding energies. 
The energy distribution curves (energy – momentum cuts) in directions  crossing these hexagons 
 has a peak at $-0.25\,$ eV and $0.45$ \AA$^{-1}$ from $\Gamma$, whose high-energy tail reaches the Fermi level (see Fig.\ref{Fig_19_B} in Appendix \ref{APPENDIX_EXP}). 
The same peak is visible at different wavevectors in the second and third BZ as a function of the photon energy and also in the first BZ at low photon energies ($\simeq$ 20 eV).  There are two tentative explanations for these states apparently well localized in energy: $i)$ they correspond to the region of the lower Hubbard band visible only in a narrow momentum range at a given photon energy because of matrix element effects of the photoemission process, $ii)$ they originate from the amorphous or defected regions. 
We tested these hypotheses analyzing a similar surface system, the B-doped Si$(111)\sqrt{3}\times \sqrt{3}$ R$30^\circ$ structure (Si-$\sqrt{3}$ in the following), which does not show the amorphous regions.

\begin{figure}
\centering
\includegraphics[width=\columnwidth]{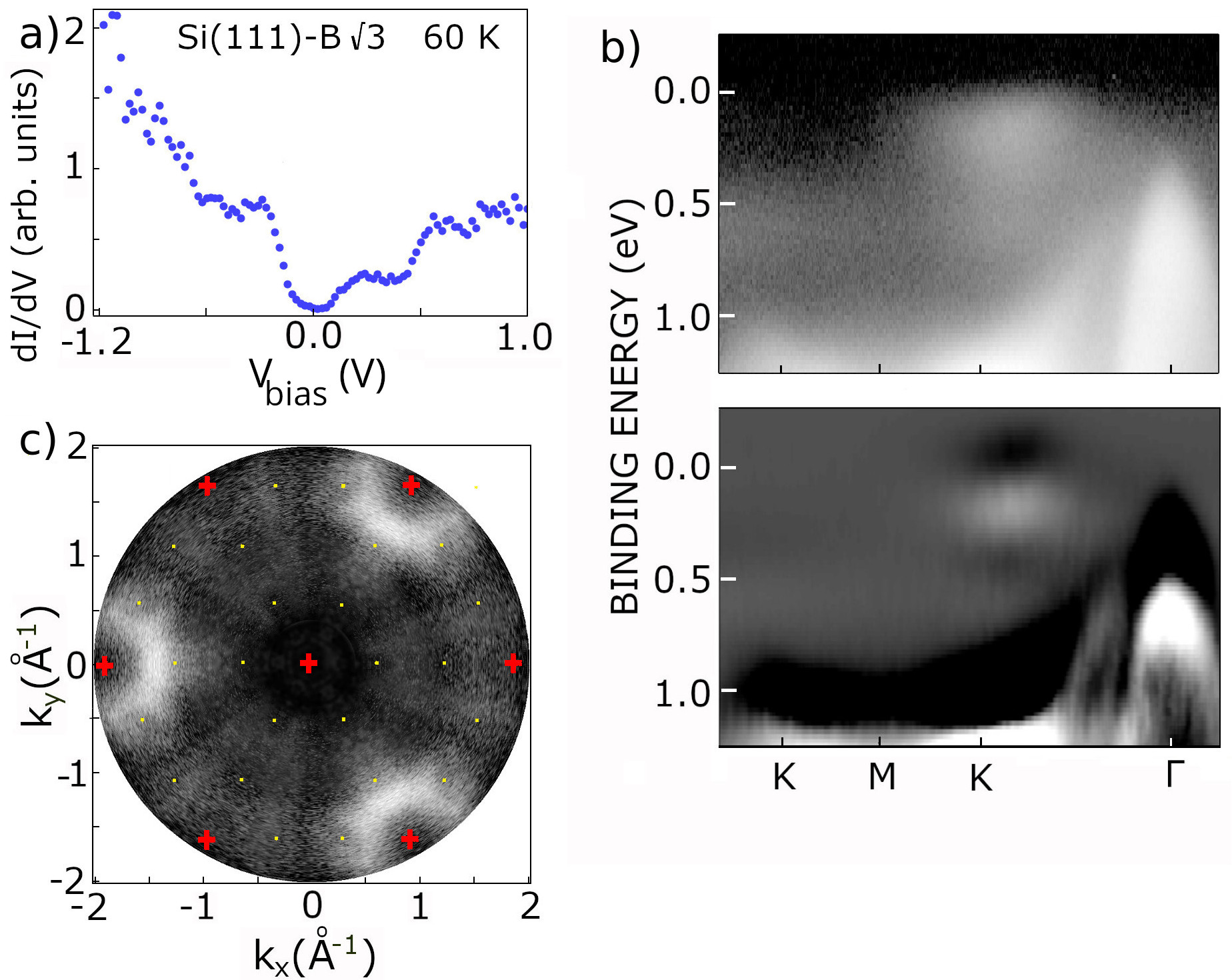}
\caption{
a) tunneling spectrum of the Si-$\sqrt{3}$ phase of Si$(111)$ stabilized by B doping with about $0.1$ ML of B at $60$ K. b) Photoemission spectral intensity along the $\Gamma K$ direction of this surface at $30$ K. The upper panel shows the raw data, the lower the opposite of the second derivative with respect to energy to reduce the background. c) map of the photoemission intensity at the Fermi level as a function of the parallel wavevector $k$. The red cosses and the yellow points are the $\Gamma$ points of the 1x1 BZs and the K poits of the $\sqrt{3}$ BZs respectively. 
}
\label{Fig_4}
\end{figure}

The Si-adatom terminated surface, stable for subsurface B concentration between less than $0.1$ and $0.33$ ML, has the same structure, the same lattice parameter, and the same number and type of valence electrons than the doped Sn-adatom terminated Sn-$\sqrt{3}$ surface. The differences are in the height of the adatoms from the substate, in the hybridization between the adatom and substrate states and in the slightly larger inter-site Coulomb repulsion $U$ on the adatoms. According to the calculation by Hansmann et al. \cite{Hansmann_2013} the hypothetical undoped Si-$\sqrt{3}$ surface with one electron in the surface band should be a Mott-Hubbard insulator at low temperature with a Hubbard gap of about 0.2 eV, slightly larger than that of Sn-$\sqrt{3}$, with a similar transition temperature, and less affected by charge fluctuations than the Sn covered surface. However, the undoped Si-$\sqrt{3}$ surface is unstable and cannot be measured. The fully B-saturated Si-$\sqrt{3}$ surface with one B atom for each Si adatom ($0.33$ ML) and an even number of electrons per unit cell is a stable band insulator with a bangap of about 1 eV and no states at the Fermi level in agreement with LDA calculations \cite{Aldahhak_2021, Shi_2002, Grehk_1992} (see below).  The same surface with a lower B concentration (less than 0.33 ML) and therefore with an average occupation number of the surface band between 0 and 1 should be instead a doped Mott-Hubbard system.
We thus measured the STS spectra on the B-doped Si-$\sqrt{3}$ surface at intermediate B concentration ($0.1$ ML) and at $60$ K revealing the existence of the Hubbard bands also in this phase (see Fig.\ref{Fig_4}a), $i.e.$ two peaks below and above E$_F$ at $-0.3$ eV and $+0.2$ eV, and very small differential conductance at E$_F$, in agreement with what is observed at room temperature on unsaturated surfaces \cite{lyo_1989, Eom_2015} and in contrast to the $1$ eV gap measured on the B saturated surface \cite{Makoudi_2017}. The low-energy peak at $-0.3$ eV is present in the photoemission data by Grehk et al. \cite{Grehk_1992} when the B coverage is below saturation and vanishes at saturation. 
The additional feature measured at about $0.6$ eV corresponds to the signal measured by STS near full doping on isolated Si adatoms without a B atom below them by Makoudi et al. \cite{Makoudi_2017, chimici, Eom_2015}. 
Our spectra show an electronic structure similar to the narrow gap Mott Hubbard state, suggesting that the doped Si-$\sqrt{3}$ surface has the same kind of physics as the doped Sn-$\sqrt{3}$.

ARPES measurements on the B-doped Si-$\sqrt{3}$ (see Fig.\ref{Fig_4}b) 
reveal both similarities and differences with respect to the B-doped Sn-$\sqrt{3}$ surface. 
We measure a maximum of photoemission intensity at about $-0.25$ eV and at 0.45 \AA $^{-1}$ from $\Gamma$ of the third BZ, similarly to the Sn-terminated case (see Fig.\ref{Fig_2}b), but we do not find any evident partially-filled band crossing the Fermi level (see appendix B for more details). As a consequence,  the map of the photoemission intensity at $E_F$ ( Fig.\ref{Fig_4}c) lacks the hexagons around the $K$ points, while it still shows features at the edges of the third BZs (resembling those seen with Sn and indicated as green hexagons in Fig.\ref{Fig_3}) which however simply originate from the high-energy tail of the $-0.25$ eV peak. 
Contrary to what observed in the case of Sn-$\sqrt{3}$, the Si-$\sqrt{3}$ surface does not show  amorphous regions at intermediate B concentration,  thus we attribute the $-0.25$ eV peak (and its related tail at the Fermi level) to the ordered Si-$\sqrt{3}$ structure and specifically to the lower Hubbard band. 
Considering that  $-0.25$ eV peak on the Si-terminated surface strongly resembles the one at $-0.3$ eV observed in Sn-terminated,  
we can attribute the large hexagons at $E_F$ (Fig.3) to the tails of the Hubbard band also in the Sn case.

In conclusion, we can consider only the small hexagons centered at K in doped Sn-$\sqrt{3}$ as doping induced Fermi surface, whose  filling (estimated from the FS's area),  is approximately $0.40\pm0.08$ electrons per unit cell, due to the Luttinger theorem.



At this point it is interesting to compare  the dispersion of the partially filled band that we observe on B-doped Sn-$\sqrt{3}$ with the quasiparticle interference results obtained by Ming et al.\cite{Ming_2017}.
The authors of Ref.\cite{Ming_2017} find a Fermi level crossing at $0.4$ \AA$^{-1}$ along $\Gamma K$ and at $0.57$ \AA$^{-1}$ along $\Gamma M$, which are indeed confirmed by our ARPES measurements. However, we cannot account for the additional reported $0.2$ \AA$^{-1}$ scattering wavevectors  along the $\Gamma$ to $M$. Indeed, according to the analysis of Ref. \cite{Ming_nature22}, the feature at $0.2$ \AA$^{-1}$ is likely a multiple-scattering effect and not a real band crossing.

The similarity between the Fermi surface we measure by photoemission on Sn-$\sqrt{3}$ and the one that was measured on the homogeneous phase by Ref. \cite{Ming_2017} indicates that the Fermi region of our spectra is dominated by contributions from the very same homogeneous phase, even if it covers about 10-20\% of the surface. This is consistent with the low intensity of the sharp metallic band. It is likely that the dishomogeneous phase (Fig.1d) that cover most of the surface contributes with broader features lost in the background. The low intensity of the Si valence band at $E_F$ suggests that these states too are associated to the minority homogeneous phase.



\section{Theoretical calculations}



From the experimental investigations, we have the following evidences for the B doped Sn-$\sqrt{3}$ system: the STM maps indicate the presence of both homogeneous/disordered Sn-$\sqrt{3}$ and amorphous regions, presenting metallic and insulating electronic phases, respectively, as measured by STS (see Fig.\ref{Fig_2}).
By ARPES measurement we discovered an electron-like band crossing the Fermi level at the K-point which likely comes from the homogeneous metallic phase and an additional feature around $\Gamma$ attributed to the tail of the lower Hubbard band.
Given this rich experimental scenario and the complicated relationship between Mott physics and doping effects, we developed different theoretical models to account for the experimental evidences.\\ 
Because of the strong correlations, largely confirmed by previous studies \cite{Profeta_2007,Modesti_2007,Biderang_2022,Li_2011,Hansmann_2013,Schuwalow_2010,Wolf_2022}, we studied the evolution of the band structure of Sn-$\sqrt{3}$ as a function of the B content, considering strong correlations effects employing both mean-field like Density Functional Theory (DFT) with local electronic correlations (DFT+U) and many-body technique within Cluster Perturbation Theory (CPT) (see Appendix \ref{APPENDIX_THEO} for theoretical details).

\subsection{Homogeneous doping effects}
\label{DFT_calculation}

\begin{figure}
\centering
\includegraphics[width=0.95\columnwidth]{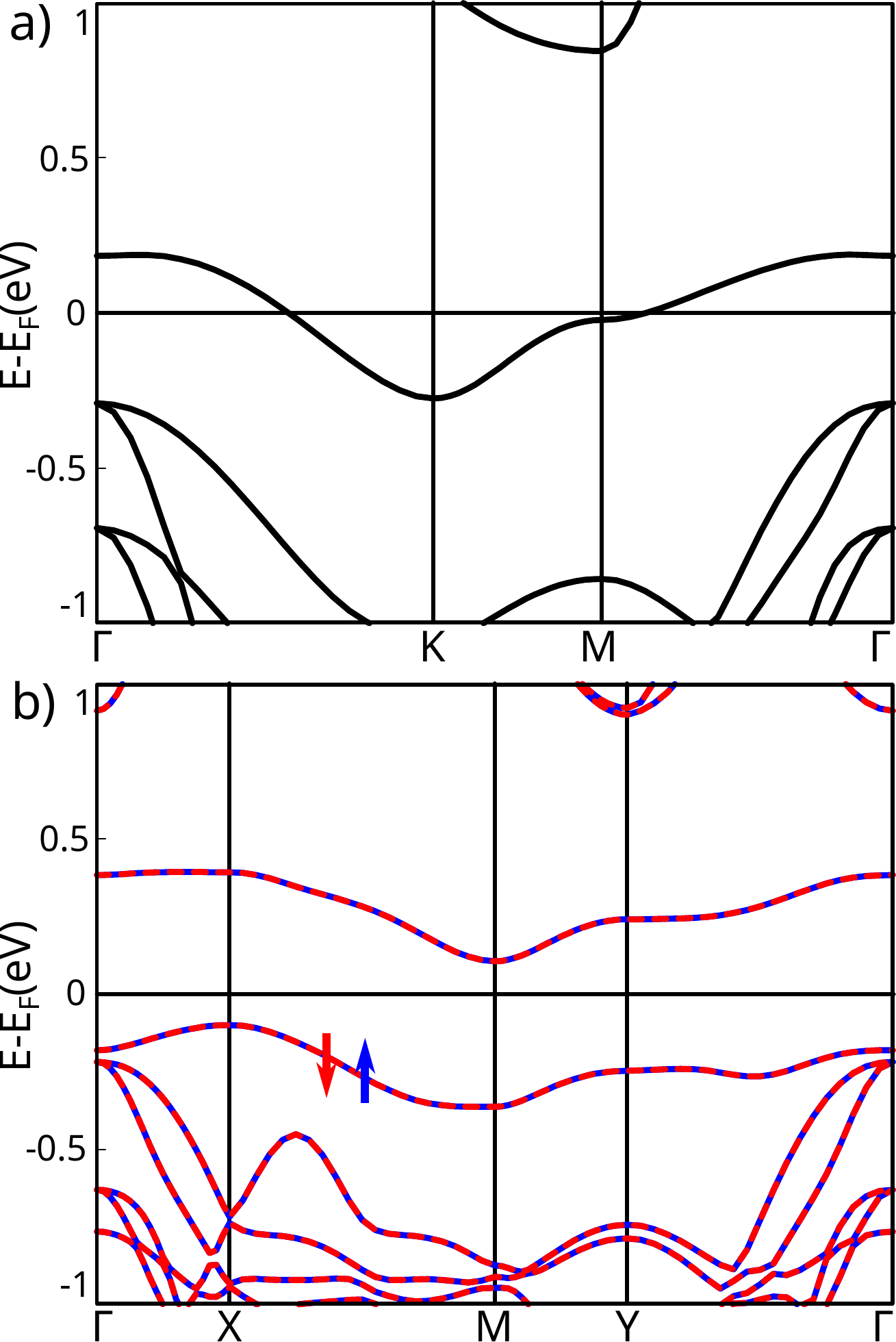}
\caption{
a)\& b) DFT+U calculated band structure in the non-magnetic and antiferromagnetic phases respectively. Red and blue lines in b) correspond to spin down (red arrow) and spin up (blue arrow) bands respectively.
}
\label{DFT_undoped}
\end{figure}

We start the discussion, reporting the electronic band structures for the ideal undoped Sn-$\sqrt{3}$ system. Disregarding magnetism (see Fig.\ref{DFT_undoped}) the system is predicted metallic with a single half-filled band crossing the Fermi energy. 
 Inclusion of magnetism at the DFT+U level, the ground state of the system is an antiferromagnetic (AFM) Mott insulator \cite{Profeta_2007}, as experimentally confirmed below $\sim 70$ K\cite{Modesti_2007}, with a possible development of a collinear $2\sqrt{3}\times\sqrt{3}$ phase \cite{Lee_2014,Li_2013}, confirmed by our calculating, whose band structure is reported in Fig.\ref{DFT_undoped}b.
During the past years, the origin of the insulating phase in Sn-$\sqrt{3}$ was indeed debated between the Mott-physics driven by strong on-site repulsion on the Sn 5p$_z$ dangling bond treated with the GGA+U framework and Slater-type insulating phase driven by magnetic exchange coupling between resonant Sn 5p$_{z}$-Si p hybrid state described with HSE06 hybrid functional \cite{Zahedifar_2019,Profeta_2007}.
In order to have a complete description of the physics of systems, we performed calculation using both methods (where possible) finding consistent results in terms of band structure and electronic properties. We report the HSE06 hybrid functional calculations of the same magnetic configurations presented in Fig.\ref{DFT_undoped} in Appendix \ref{APPENDIX_THEO} for comparison.
Despite being two different approaches, both HSE06 and DFT+U capture the main physics of the system for both doped and undoped phases (see below). Thus, in the following we present only DFT+U results, while HSE06 ones are contained in the Appendix \ref{APPENDIX_THEO}.


Considering the experimental evidences, which show the coexistence of metallic and Hubbard bands, we investigate the doping effects of B atoms within a rigid band approximation.
Including an hole doping of $\Theta_B=10\%$ (corresponding to hole doping at which the superconducting phase is observed \cite{Ming_2017,Ming_2023}), the non-magnetic band is rigidly shifted at higher energies, emptying the van-Hove singularity peak at the M point. 
Although DFT accounts for the metallic band at K, it lacks the presence of correlated Hubbard bands observed in the experiments.
Thus, in order compare with experimental results, we need to include many-body effects in the doped correlated system.
We thus employ CPT (see Appendix \ref{APPENDIX_THEO})


\begin{figure}
\centering
\includegraphics[width=\columnwidth]{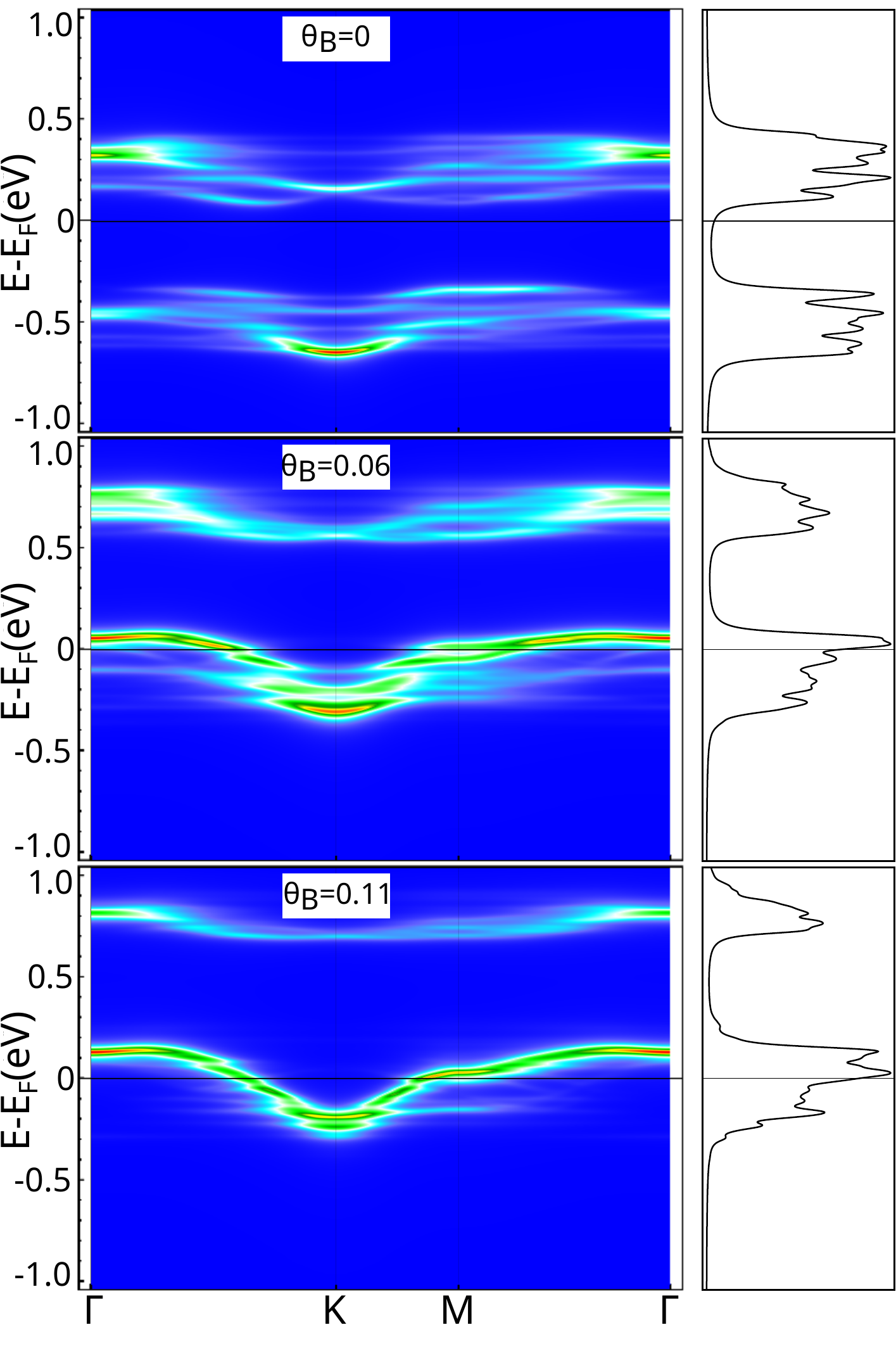}
\caption{
 Calculated spectral function for no ($\Theta_B=0$) one ($\Theta_B=0.06$) and two ($\Theta_B=0.11$) hole doping along with the corresponding DOS.
}
\label{CPT}
\end{figure}

CPT calculations were performed on the undoped MI surface band of the Sn-$\sqrt{3}$ phase (Fig.\ref{DFT_undoped}a). Doping is then simulated by integer number of holes  within the $6$-Sn cluster used (see Fig.\ref{cluster_green}). A value of the Hubbard $U\simeq 0.7$ eV is assumed to reproduce a MI gap of $\sim 0.3 \divisionsymbol 0.4$ eV, as in DFT calculations (see Fig.\ref{DFT_undoped} and \ref{DFT_undoped_APPENDIX}).\\ 
Fig.\ref{CPT} shows the one particle spectral function $A\left(\boldsymbol{k},\omega \right)$ as a function of $\boldsymbol{k}$, along the Sn-$\sqrt{3}$ symmetry points, and of the energy $\omega$ for the DOS. 
In the undoped phase, we can see the clear MI gap with lower and upper Hubbard bands. Increasing the doping level at $\Theta_B\simeq 5\%$, the system becomes a metal still retaining, in part, the Hubbard bands and partially developing a quasi-particle peak at the Fermi level, resembling the experimental ARPES spectrum (see Fig.\ref{Fig_3}).
This result is in complete agreement with DMFT results obtained in Ref.\cite{Cao_2018} at similar doping levels confirming that the appearance of a QPP between the lower and upper Hubbard bands has to be attributed to a spectral weight transfer that is a purely many body phenomena. \\
As the doping is further increased, the spectral function of the UHB, although still present, loses spectral weight indicating a suppression of the magnetic order and the restoring of a quasi-particle picture.
The Fermi wavevector does not change appreciably when the doping is increased by a factor two, in agreement with the experimental results of Fig.2c. 
From the above analysis we note that the strong-coupling regime is clearly evident at intermediate doping levels, in which our spectrum shows the best agreement with the ARPES and STS results and in line with the doping level at which superconductivity seems to develop \cite{Wu_2020,Ming_2023}. However, at this doping regime, the STM maps indicate that most of the surface is not in the homogeneous phase, as B doping induces a rather apparent dis-homogeneous corrugated Sn layer, as in the B-doped Si-$\sqrt{3}$ analog, as discussed in Section \ref{Experimental}. Moreover, the full doped homogeneous surface (with 1 B per Sn), is experimentally found metallic (as we have found and in agreement with the results of Ming et. al.\cite{Ming_2017,Wu_2020}), instead of an insulating phase expected from the discussed rigid-band scenario. 
We thus conclude that, despite the apparent good agreement of the calculated spectral function in Fig.\ref{CPT} with the ARPES spectra in Fig.\ref{Fig_2}, in order to have a complete understanding of the physics the doped phases, an in-depth analysis of chemical B doping is crucial.

\subsection{DFT description of doped phases}

\begin{figure}
\centering
\includegraphics[width=\columnwidth]{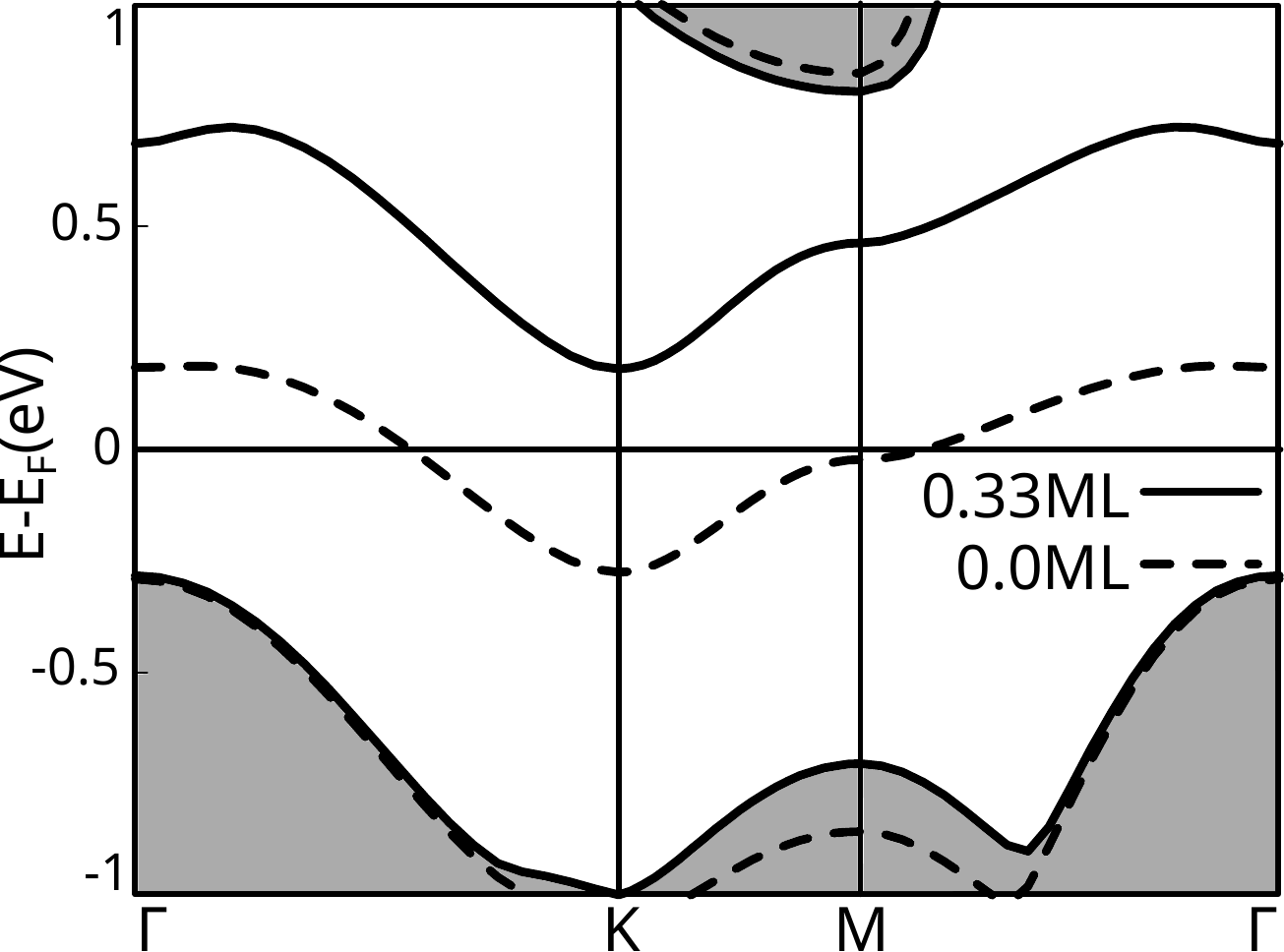}
\caption{
 Non-magnetic band structure of the fully doped ($\Theta_B$=0.33 ML) surface (red) compared with the one of the undoped surface (black).
}
\label{DFT_doped_B_r3}
\end{figure}

We first analyze the chemical B-Si substitution by first principle calculations, considering a B atom in the S$_5$ site (the site just below the Sn T$_4$ position, see Fig.\ref{fig1new}) at saturation doping ($\Theta_B$=0.33 ML) in the Sn-$\sqrt{3}$ unit cell. The calculated band structure of such a state are reported in Fig.\ref{DFT_doped_B_r3}. Once the B atoms substitute the Si in S$_5$ with one electron less, Sn adatom relaxes getting closer to the silicon substrate of about $\simeq 0.34$ \AA . This distortion comes together with the raising of the Sn surface band (see Fig.\ref{DFT_doped_B_r3}) which is completely emptied, resulting in an insulating electronic phase.

\begin{figure}
\centering
\includegraphics[width=\columnwidth]{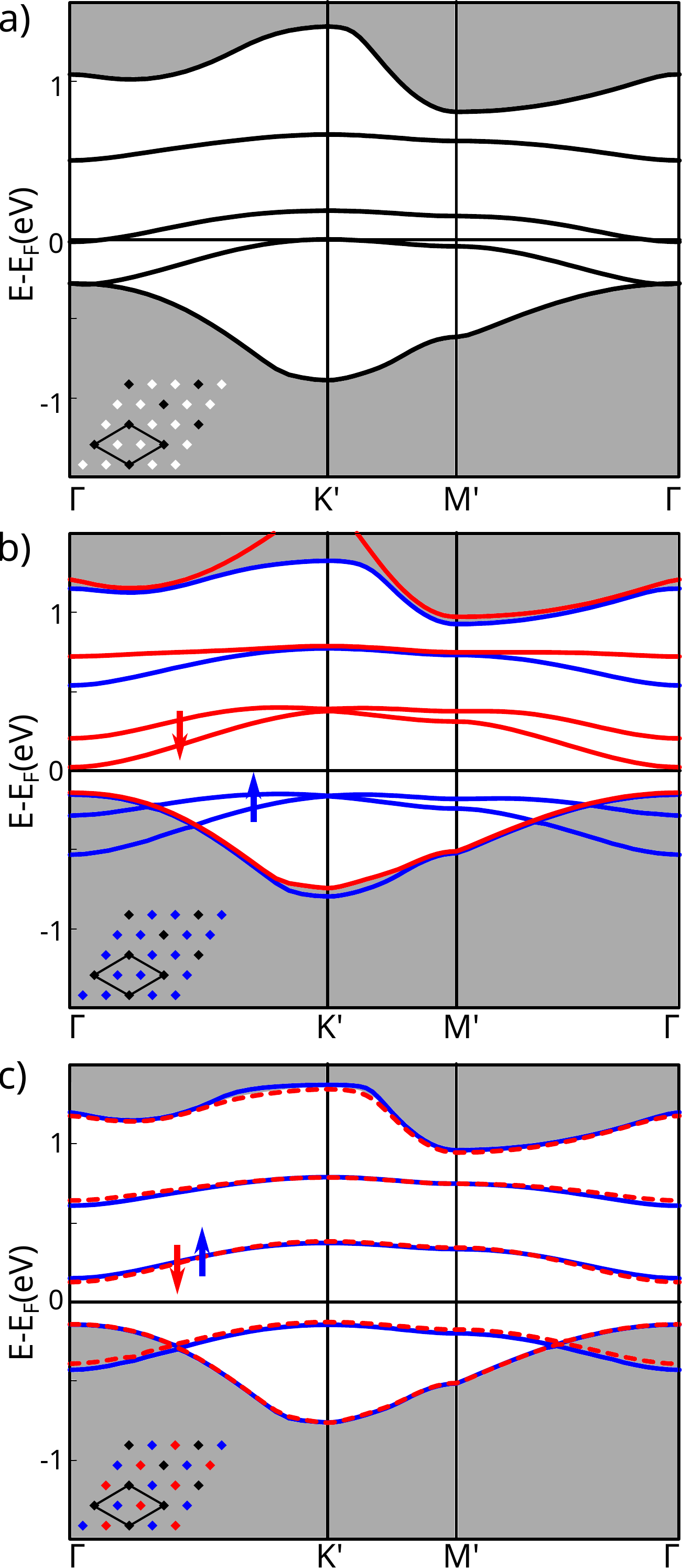}
\caption{
a), b) and c) band structure calculations, in the non-magnetic, ferromagnetic and antiferromagnetic $3\times 3$ supercells (black rhombus in the insets) respectively with only one boron ($\Theta_B$=0.11 ML) in the S$_5$ position, done with PBE+U functional. Red and blue lines in b) and c) correspond to spin down (red arrow) and spin up (blue arrow) bands respectively. The insets are sketch of the structure used in the respective band structure calculation: black, white, blue and red circles represent B-doped, non-magnetic, spin up and spin down Sn adatoms.
}
\label{DFT_doped_B}
\end{figure}

However, the experimental doping levels discussed in the previous sections are lower than 0.33 ML. We thus consider a model system to describe a doping closer to the experimental ones,  $\sim 0.1$ ML B coverage (one B over 3 Sn adatoms) using a $3 \times 3$ supercell considering the non-magnetic, ferromagnetic and collinear antiferromagnetic phases (which are possible because of the complete emptying of the Sn dangling orbital) and report the corresponding band structures in  Figure \ref{DFT_doped_B}a-c). \\

Interestingly, similarly to the fully doped surface, even at fractional doping, the trivalent B atom induces a complete capture of the Sn p$_z$  electron with the hole localized on the Sn adatom, which then appears dark in filled states STM topography. 
This produces a nearly insulating phase even  when we consider the non-magnetic configuration (we find a small overlap between two bands,see panel (a) in Fig.\ref{DFT_doped_B} whose origin is discussed in \ref{APPENDIX_THEO} also using different functionals), while the lowest energy magnetic configurations are insulating \footnote{The total energy difference between antiferromagnetic and ferromagnetic configurations is of 13 meV and 14 meV for DFT+U and HSE06 calculation respectively, favoring the antiferromagnetic phase.}, rather than metallic as naively expected from the shifting the Fermi energy upon hole doping.
This electronic effect is accompanied  by a structural distortion of  of the Sn adatom  with B underneath  towards the substrate by 0.33 \AA , which should be the origin of the 
 topographic disorder observed in some regions of the STM maps (see Fig.\ref{fig1new}d) where dark (bright) spots are correctly interpreted as Sn adatoms with (without) a B atom underneath. 
Thus, even considering different B-doping levels, the hole-localization mechanism occurs, eventually leading to an insulating phase, at variance with the experimental STS and ARPES spectra presented above, which  instead show metallic phases of non-homogeneous regions.

On the other hand, the metallic homogeneous regions cannot be explained by a model which consider B atoms in S$_5$, that brings to a corrugated and a band-insulator phase. These considerations
point to a further refinement of the theoretical models to account for the experimental evidences in both non-homogeneous and homogeneous  regions of the doped surfaces.

\subsubsection{Doping-dependent electronic screening}\label{screened hubbard}

DFT+U approximation requires the use of the $U$ parameter to accounts for the on-site repulsive correlation effects, which is strictly related to the chemical environment of Sn adatoms \cite{Cococcioni_2005} and it should be affected by the Si-B substitution. Thus, we re-calculated the U parameter for the Sn adatoms in the  doped $3\times 3$ supercell, indeed finding different values with respect to the undoped one. We get $U_{\text{Sn/B}}\simeq1.5$ eV and $U_{\text{Sn/Si}}\simeq2.2$ eV for the Sn with or without the B dopant in the S$_5$ position, respectively. 

\begin{figure}
\centering
\includegraphics[width=\columnwidth]{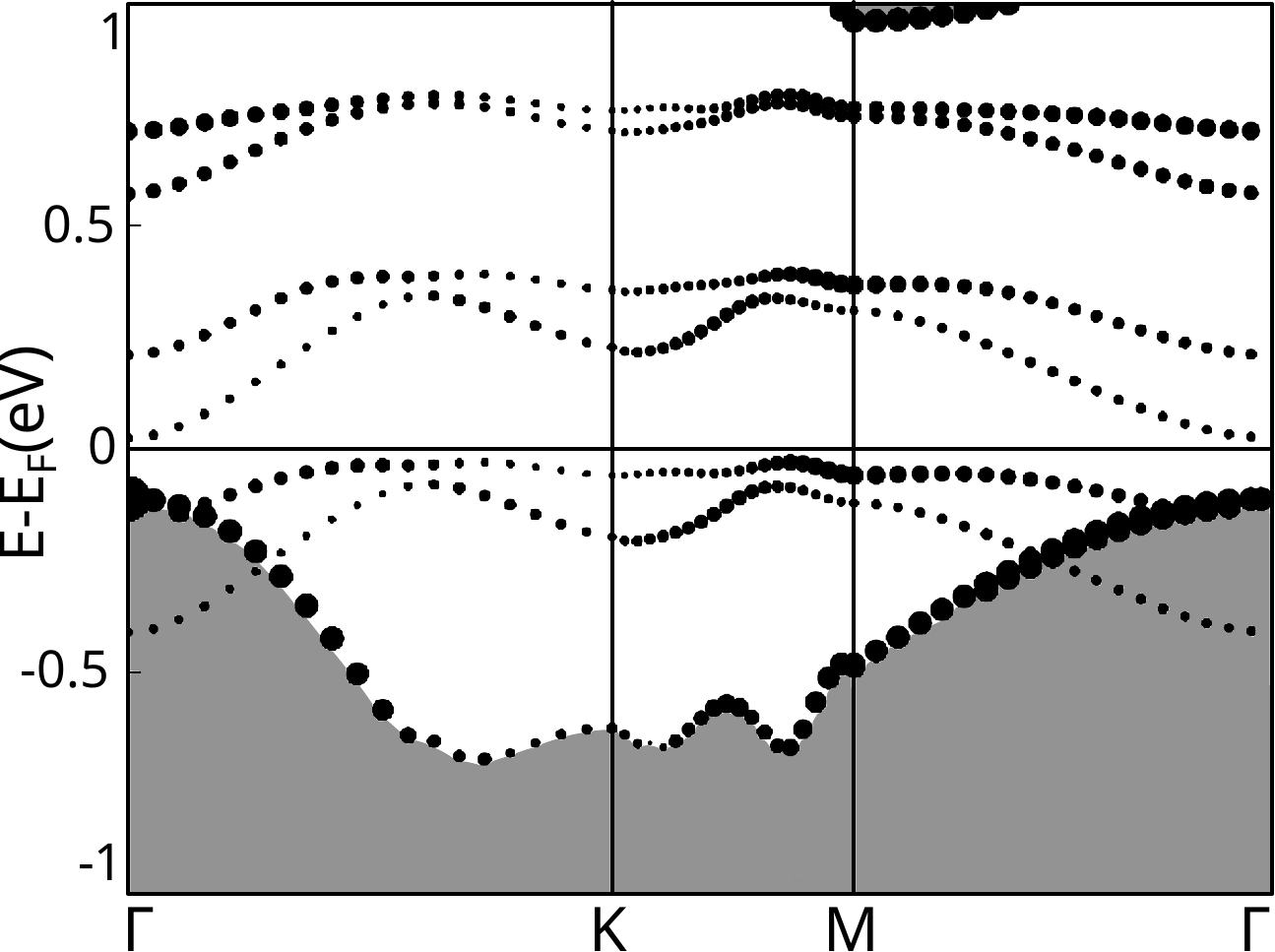}
\caption{
Image reporting the $3\times 3$ supercell band structure calculated in DFT+U with one B atom in the S$_5$ position ($\sim 0.1$ ML B coverage) with re-computed U parameter due to the change of the chemical environment (see text). The band structure is unfolded in the $\sqrt{3}\times\sqrt{3}$ BZ. 
}
\label{DFT_doped_B_U_changed}
\end{figure}

The calculated band structure using these new values in the AFM phase (see Fig.\ref{DFT_doped_B_U_changed}) still shows insulating phase but with a strongly reduced Mott gap ( $\sim 0.06$ eV). 

The increase of the screening due to the B doping, thus produce a tendency to a metallic phase, even if the effect is still not strong enough to account for the experimental metallicity of the doped systems. 

However, smaller $U$ corrections could be expected in the real surface considering multiple effects: $i)$ The real Boron sublattice is expected to be disordered and possibly giving rise to larger screening on Sn adatoms, which cannot be properly described by any periodic models, like the $3\times 3$ supercell used; $ii)$ The linear response method \cite{Cococcioni_2005} that we are using to calculate $U$ may not be accurate in describing high occupation changes in a strong coupling regime as the hole-localization.\\
These considerations point towards the inclusion of disorder effects to describe the boron doping, which could explain the coexistence of MI and metallic states as a function of the B concentration observed in non-homogeneous regions. 

On the other hand, the origin of the metallicity of the  well-ordered (homogeneous) sections of the Sn-$\sqrt{3}$ surface, at fractional doping is yet to be explained.

\subsubsection{Sub-surface doping}

As suggested by the experiments in Ref. \cite{Makoudi_2017,Ming_2023} and anticipated in Section \ref{Experimental}, it can be possible that B atoms are not exclusively in the S$_5$ site. 
We thus assume that, in ordered domains, the B atoms should be buried deeper in the surface producing a rather homogeneous STM topography.

\begin{figure}
\centering
\includegraphics[width=\columnwidth]{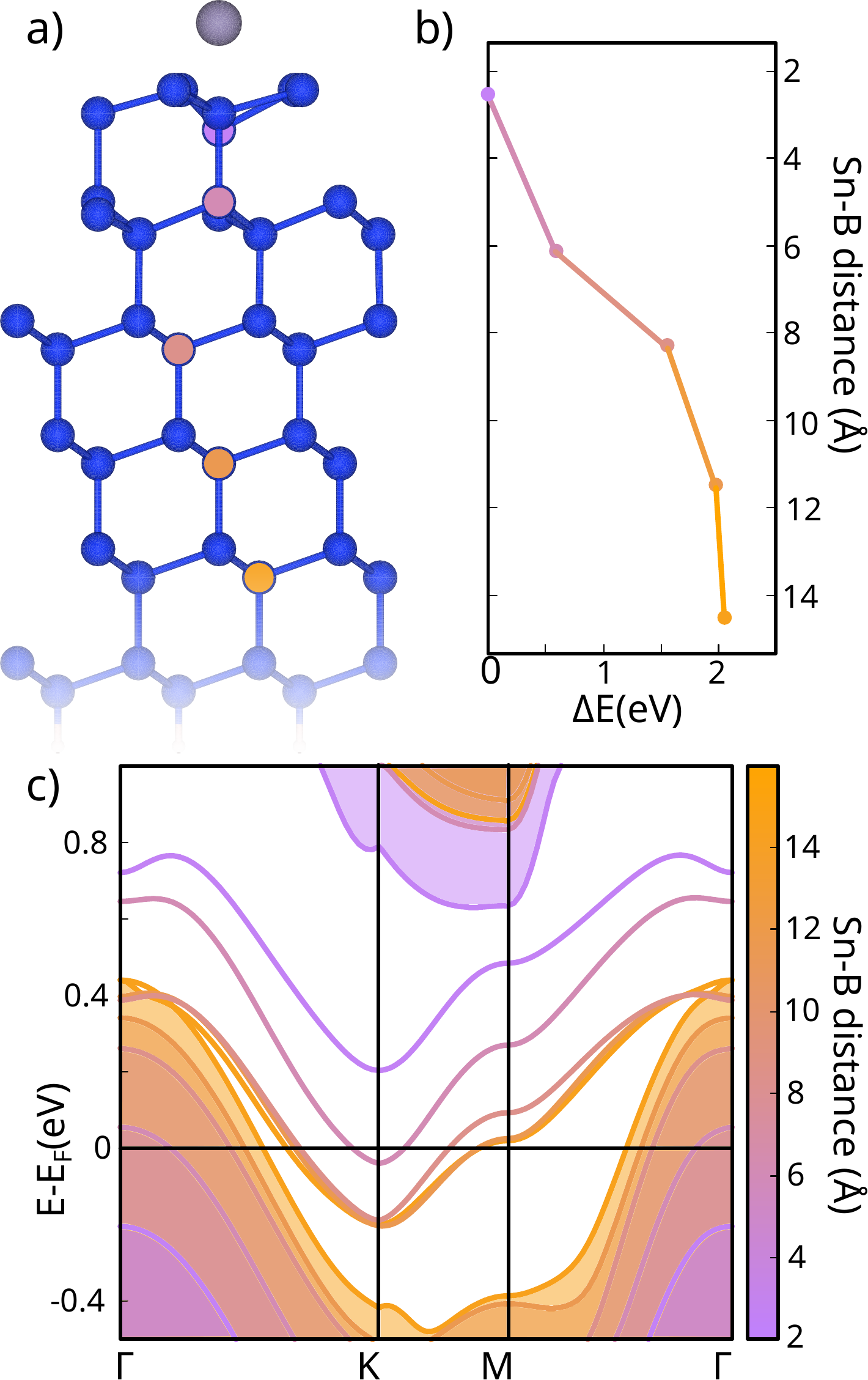}
\caption{a) The $6$ Si Bi-layer structure used in non-magnetic DFT+U calculation. b) the total energy difference per B atom $\Delta E$, with respect to the one calculated with the B atom in the S$_5$ site, as a function of the Sn-B distance. Sn, Si, B and H atoms are represented as grey, blue, green and white spheres respectively. c) The computed band structure of the construction in a) for various Sn-B distances (colormap).
}
\label{DFT_pbe_6Bi}
\end{figure}

In Fig.\ref{DFT_pbe_6Bi}a we report the total energy difference \footnote{Note: we used a $\sqrt{3}\times \sqrt{3}$ unit cell.} $\Delta E=E_{B_i}-E_{B_{S_5}}$ where $E_{B_i}$ ($E_{B_{S_5}}$) is the total energy of the structure in which the B atom is in the $i$-th bi-layer (S$_5$) position (see Fig.\ref{DFT_pbe_6Bi}). 
From Fig.\ref{DFT_pbe_6Bi}b we evince that the S$_5$ site is clearly the most stable for boron substitution, while the other sites are  metastable sites where B atoms can be trapped.


The band structure relative to the systems with B in different doping sites, is reported in Fig.\ref{DFT_pbe_6Bi}c. 
As discussed before, when B is in the S$_5$ site, the Sn derived surface band is empty and well above the Fermi energy, while already when B replaces silicon in the second bi-layer, the ionization of the Sn dangling bond is not complete, resulting in a partially occupied surface band at the K-point.
The effect is even more evident for deeper sites.
The bulk-like B atom (Sn-B distance $\gtrsim 12$ \AA) mostly hybridizes with bulk Si orbitals, shifting the Fermi energy right into the Si bands that eventually will form the acceptor levels in the bulk.
When the B is in the second bilayer the top of the calculated Si valence band touches the Fermi level, in agreement with the experimental results of Fig.\ref{banaval.jpg}d. 
This partial doping mechanism can explain the metallization of the homogeneous Sn-$\sqrt{3}$ surface, $de-facto$ resembling the rigid-band picture discussed in Sect.\ref{DFT_calculation}, with the van Hove singularity just above the Fermi energy.
In fact, the Supplementary Material of Ref.\cite{Ming_2023} also attributes the counter intuitive metallicity of Sn-$\sqrt{3}$ to an incomplete ionization of B atoms near the surface and/or to a less-then-1/3 ML B concentration. This hypothesis is supported by our detailed evidence.

\subsection{Disorder effects}\label{disorder}

\begin{figure}
\centering
\includegraphics[width=\columnwidth]{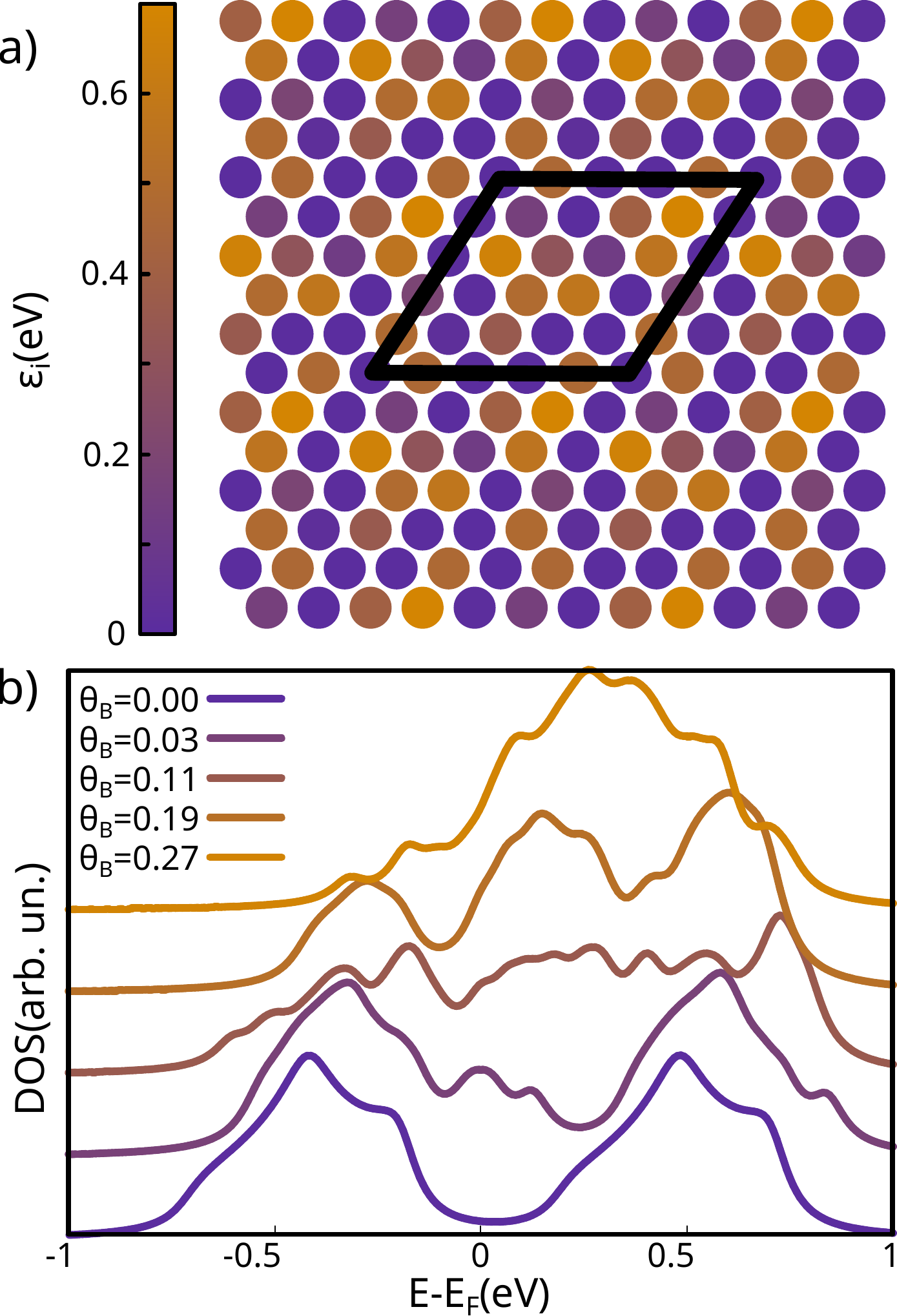}
\caption{
a) Example of one disordered $5\sqrt{3}\times 5\sqrt{3}R30^{\circ}$ configuration with sites color scale corresponding to the onsite energy. b) Density of states calculated in the $5\sqrt{3}\times 5\sqrt{3}R30^{\circ}$ supercell changing number $N_B$ of B atoms.
}
\label{TB_fig2}
\end{figure}

The theoretical models we discussed so far are all based on the assumption of a perfect periodicity, without any disorder effect taken into account. This is in contrast with some of the experimental STM topographies which show mostly inhomogeneous and disordered B dopants.
To take into account disorder effects, we model the B-doped Sn/Si$(111)$ structure by considering the following Hamiltonian:

\bea \label{TB_H}
H &=& \sum_{ij\sigma}t_{ij}\left(c^{\dagger}_{i,\sigma}c_{j,\sigma} + h.c.\right)+
\sum_{i,\sigma}\epsilon_{i}c^{\dagger}_{i,\sigma}c_{i,\sigma}+\\
&+& U\sum_{i}c^{\dagger}_{i,\uparrow}c_{i,\uparrow}c^{\dagger}_{i,\downarrow}c_{i,\downarrow}+
V\sum_{<ij>}\sum_{\sigma\sigma'}c^{\dagger}_{i,\sigma}c_{i,\sigma}c^{\dagger}_{j,\sigma'}c_{j,\sigma'},\nn
\eea

where $t_{ij}$ are the hopping amplitudes among Sn sites $i$ and $j$, $c^{\dagger}_{i,\sigma}$ is the creation operator of an electron in site $i$ and spin $\sigma$, $\epsilon_i$ is the onsite energy on site $i$ (taken at zero for the ideal undoped Sn-$\sqrt{3}$ phase), $U$ is the Hubbard parameter and $V$ is the non-local nearest-neighbor Coulomb interaction. We solve the Hamiltonian in Eq. \ref{TB_H} within the mean field theory (MFT) using a self-consistence procedure \cite{pyqula}, recently demonstrated to be a valid approach to predict the effect of local and non-local interaction, charge order, spin order and disorder \cite{Lado_2024} on 2D systems.
We simulate disordered B-doping using a $5\sqrt{3}\times  5\sqrt{3}R30^{\circ}$ supercell and considering hopping amplitudes up to the $6^{th}$ nearest neighbors as fitted from first principle band structure in the non-magnetic phase. A local Hubbard parameter of $U=0.9$ eV is assumed to reproduce a MI gap of $\simeq0.4$ eV, while the non-local interaction $V\in \left[ -0.2, 0.5 \right]$ eV
is taken as a parameter varied to explore the phase diagram. Doping and disorder effects are simulated  randomly choosing $N_B$ positions, where $N_B$ is the number of boron atoms for which the on-site energy is randomly chosen within 
$\epsilon_{i=1,\dots,N_B}\in \left[0,\epsilon_0\right]$ (see Fig.\ref{TB_fig2}) where $\epsilon_0\simeq 0.70$ eV \footnote{$\epsilon_0$ is the $ab-initio$ on-site energy of the Sn p$_z$ band of the fully B-doped surface, calculated by first-principles using PBE+U functional in the NM phase} and the number of electrons reduced by $N_B$.
Therefore, three different sources of disorder are included in our model: $i)$ structural disorder due to the random 2D location of dopants and, by using a continuous distribution of on-site parameters, $ii)$ the variation of the electronic screening on the on-site Coulomb repulsion on Sn, as discussed in Sect.\ref{screened hubbard} and $iii)$  the different degree of B-Sn hybridization caused by the depth distribution of B dopants from the surface.

\begin{figure*}
\centering
\includegraphics[width=2\columnwidth]{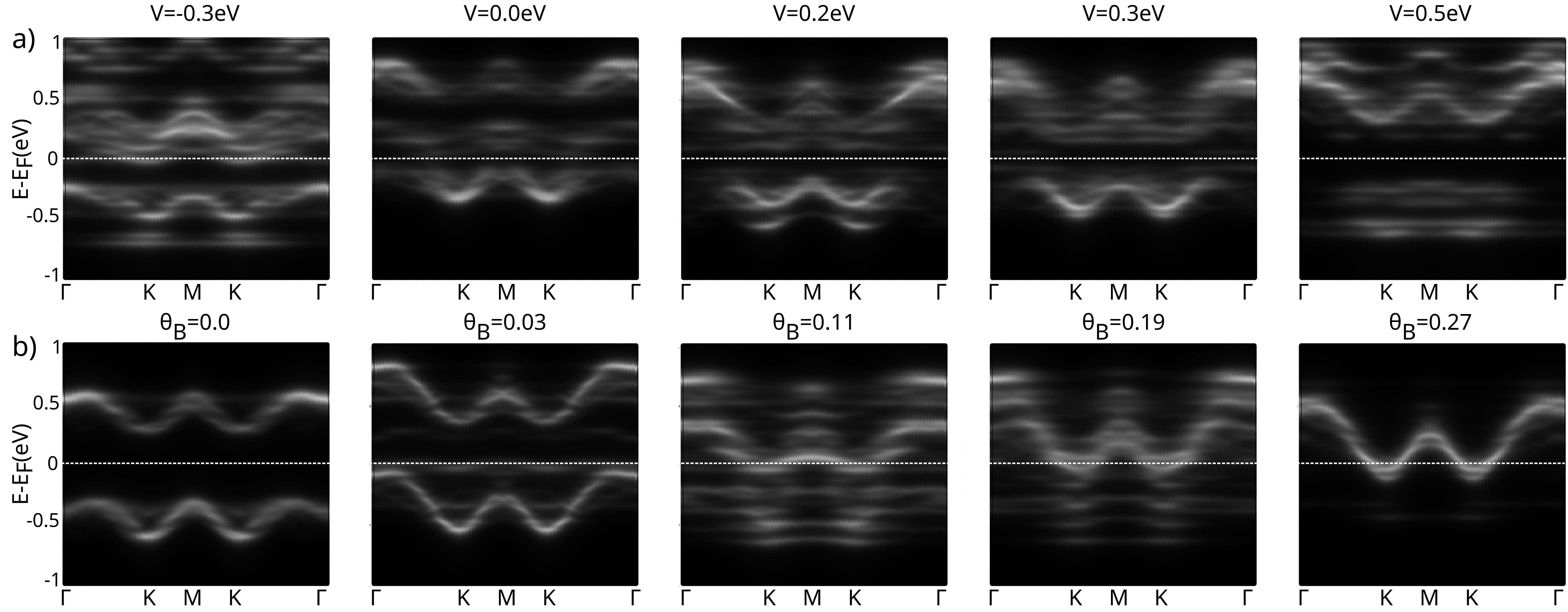}
\caption{
Calculated spectral function of a $5\sqrt{3}\times 5\sqrt{3}R30^{\circ}$ supercell along $\Gamma KMK\Gamma$. We used $U=0.9$ eV, $N_B=8$ and $V=-0.3,0,0.2,0.3,0.5$ eV in top figures (a) while $U=0.9$ eV, $V=-0.2$ eV and $N_B=0,2,8,14,20$ in bottom figures (b).
}
\label{TB_fig1}
\end{figure*}

In order to study the evolution of the electronic properties as a function of the doping level we calculated the density of states (Fig.\ref{TB_fig2} b) and the spectral function $A\left( \boldsymbol{k},\omega \right)$, which can be directly compared with the experimental ARPES spectra, Fig.\ref{TB_fig1}.

The calculation of density of states confirms that when no boron atom are present, the DOS shows two main peaks corresponding to the LHB and UHB with the Fermi energy in the middle. As $N_B$ increases, a metallic in-gap state appears acquiring spectral weight, similarly to the QPP observed experimentally (see Fig.\ref{Fig_2}). For high B concentration the LHB and UHB are no more visible with the metallic state being dominant.\\

The calculated spectral functions for different values of the non-local $V$ interaction are reported in Fig.\ref{TB_fig1}a: for positive $V$ parameter, even though in-gap states are present, the Fermi energy lies just on top of full occupied states corresponding to an insulating GS (Fig.\ref{TB_fig1}a) and to reproduce the experimental band dispersion at the Fermi energy, a negative $V$ parameter is needed. The possible origin of an attractive NN is discussed in Ref.\cite{Lado_2024} and reference therein.
In particular, for $V=-0.2$ eV, the evolution of the spectral function at various B concentrations is showed in Fig.\ref{TB_fig1}b where we find an in-gap metallic band appearing at E$_F$ and acquiring spectral weight  between the LHB and UHB around the $K$-point of the BZ. This metallic band will form, at higher doping, the NM (but rigidly doped) Sn band in Fig.\ref{DFT_undoped}a.

In conclusion, the evidences, from STS spectra, of metallic phase in the non-homogeneous regions can be attributed to disorder effects on the electronic screening of the on-site Coulomb interaction at variance with the periodic DFT calculations.

\section{ Discussion and conclusions}

In this paper we investigated the structural and electronic properties of the normal phases to understand the origin of the superconducting phase in B-doped Sn-$\sqrt{3}$, a system which after twenty years is still showing surprising physical phenomena.
We have addressed the problem  both experimentally and theoretically, unveiling the role of doping in this strongly  
correlated system. 

At the nanometer scale, STM and STS measurements show that the B doped surface presents at least three different types of domains, namely homogeneous (with low or high defect density), non-homogeneous and amorphous, associated with distinctive electronic phases (see Fig.\ref{fig1new}). Even at low doping levels (0.04 ML), the ARPES spectra show a metallic band, crossing the Fermi energy around the $K$ point, that we attribute to the Quasi-Particle Peak (QPP) observed in STS measurements in the homogeneous phase, which hosts the superconducting phase at low temperature \cite{Ming_2017,Wu_2020,Ming_2023}.
As the doping increases the metallic band near $K$ shifts down in energy with a corresponding small increase of the Fermi wavevector, coherently with an homogeneous doping of the Hubbard bands (see Fig.\ref{Fig_2}).
To rationalize this experimental evidence, we conducted extensive theoretical and computational studies using ab-initio Density Functional Theory (DFT), Cluster Perturabtion Theory (CPT) as well as model Mean Field Theory (MFT) calculations,  capturing different aspects of the physics of the hole doped Mott insulator phase.\\
Indeed, the CPT spectral function of the doped system strongly resembles that observed in ARPES, showing a QPP at the Fermi energy, and thus supporting a rigid band (delocalized) picture of the hole-doping of B atoms. 
However, the observed metallization cannot originate from the expected B substitution at the S$_5$ site (immediately below the Sn adatom, a site that is the lowest-in-energy and the one experimentally observed on clean Si surface).
In fact, S$_5$ site Si-B substitution leads, according to our DFT calculations, to hole localization and an essentially insulating state, also accompanied by a substantial lowering of the adatom at the doping site towards the silicon substrate. 
This will result in a different apparent height of Sn adatoms in STM maps at B coverage below saturation which would appear disordered (non-homogeneous) depending on the random location of B atoms. 
The homogeneous phase, showing the QPP, negligible corrugation of the adatom lattice, is then not compatible with B substituting Si {\em only} at the S$_5$ site.\\
Searching for a rationalization of these controversial evidences we concluded that at least some B atoms must be buried deeper in the surface. In this case, the tin dangling bond would not be subject to a full electronic capture by means of boron atoms, resulting from the overlap with the Sn p$_z$ orbital, thus giving rise to the effective rigid-band doping of the Sn surface state (Fig.\ref{DFT_pbe_6Bi}), a flatter surface and a slight depletion of the valence band, as experimentally observed. The mechanism that can give rise to regions with buried B layers is the Si epitaxial growth on the Si(111)-B substrate caused by the Si adatoms that are dislodged by the Sn atoms during the growth of the Sn overlayer. At the temperature at which the ordered Sn layer is formed, the diffusion and surface segregation of B is low and most of the dopants remain buried in the second bilayer. Core level photoemission from the B1s level support this picture with an appreciable attenuation of the B signal when Sn adatoms substitute the Si adatoms. According to this model, the main defects observed on the homogeneous phase, i.e. single adatoms deeper than the others (see Fig 1b and c), can be attributed to a limited B surface segregation \cite{Zotov_1996, Headrick_1990,Korobtsov,Schulze} that bring some of the dopants in the T$_5$ site again. The Sn atoms on top of these dopants should be about 0.3 \AA\ below the others (see Section III B), in qualitative agreement with the STM images.
\footnote{Regions with inserted Si bilayer that form before others have a longer time to allow the surface segregation of some B. This can explain the large variation of defect density on the surface.} 
We further extended our theoretical interpretation modeling the inhomogeneous phase. Based on the above considerations, these regions should originate from {\em random} B substitution on the sub-surface S$_5$ site, leading to a corrugated metallic phase induced by disorder effects on the electronic screening (Fig.\ref{TB_fig1}), in agreement with STS measurements (Fig.\ref{fig1new}). This phase forms in the regions of the sample not covered by the dislodged Si adatoms.
In conclusion, our results indicate that homogeneous regions of the Sn-$\sqrt{3}$ doped phase, in which superconductivity has been measured \cite{Ming_2023,Ming_2017}, is most likely the result of the burial of some section of the Si-$\sqrt{3}$ doped surface during the deposition process of Sn atoms. 
Our results clarify the metallization mechanism of the 2D Mott Insulating phase of Sn-$\sqrt{3}$ indicating the crucial role of deep Boron dopants. \\ 
A more refined growth process in this and in similar systems, for istance by molecular beam epitaxy, can reasonably allow a better control of the dopant-adlayer distance and therefore a finer tuning of the occupation of the surface states, of the screening, and in general of electronic properties of the surface. The rich phase diagrams of many tetravalent adlayers on Si(111) will have a new control knob to reach new intriguing states.\\
In addition, we have shown how disorder in the dopant distribution can also lead to metallicity, suggesting a possible exploration of their superconducting properties, unexplored so far.


\section*{Acknowledgments}

C. T. acknowledges financial support under the National Recovery and Resilience Plan (NRRP), Mission 4, Component 2, Investment 1.1, 
funded by the European Union – NextGenerationEU– Project Title “DARk-mattEr DEVIces for Low energy detection - DAREDEVIL” – CUP D53D23002960001 - Grant Assignment Decree No. 104  adopted on 02-02-2022 by the Italian Ministry of Ministry of University and Research (MUR) and Università degli Studi di Perugia and MUR for support within the project Vitality.\\
C. T. and T. C.  acknowledge financial support under the National Recovery and Resilience Plan (NRRP), Mission 4, Component 2, Investment 1.1, funded by the European Union – NextGenerationEU– Project Title "Symmetry-broken HEterostructurEs for Photovoltaic applications - SHEEP" – CUP B53D23028580001 - Grant Assignment Decree No. 1409  adopted on 14-09-2022 by the Italian Ministry of Ministry of University and Research (MUR).\\
C. T and G.P. acknowledge support from CINECA Supercomputing Center through the ISCRA project and Laboratori Nazionali del Gran Sasso for computational resources.\\
G.P. acknowledges financial support by the European Union – NextGenerationEU, Project code PE0000021 - CUP B53C22004060006 - “SUPERMOL”, “Network 4 Energy Sustainable Transition – NEST”.\\
Research at SPIN-CNR has been funded by the European Union - NextGenerationEU under the Italian Ministry of University and Research (MUR) National Innovation Ecosystem grant ECS00000041 - VITALITY.\\
P.M.S. and P.M. acknowledge EUROFEL-ROADMAP ESFRI of the Italian Ministry of University and Research and acknowledge Elettra Sincrotrone Trieste for providing access to its synchrotron radiation facilities (Proposal nr 20220464, 20225074, 20235246).\\

\appendix
\section{Theoretical methods} 
\label{APPENDIX_THEO}

In DFT calculations we modelled the system using 3 Si bi-layer along the [111] direction, terminated with hydrogen atoms saturating  the Si dangling at bottom of the slab.  We used the experimental silicon lattice constant of a=5.43 \AA\ and a vacuum region of at least $10$ \AA\ (see Fig.\ref{DFT_undoped}a). H atoms and the last Si bi-layer are fixed during calculations with the H-Si distance at $\sim 1.50$ \AA , obtained  through the hydrogen relaxation on the clean Si(111) surface.
We used the PBE \cite{PBE_1996,PBE_2008} exchange-correlation potential and, in order to account for strong correlation effects on Sn pz orbital, we included the Hubbard correction with first-principle calculation. The $U$ parameter is estimated by the linear response technique described by Cococcioni et al. \cite{Cococcioni_2005} resulting in $U\simeq 3$ eV. Calculation with DFT+U were done using the QuantumEspresso code \cite{QE_2009,QE_2017,QE_2020} with a converged energy cut-off of $42$ Ry on the wavefunctions and of $420$ Ry on the charge density. A $k$-space integration grid of $12\times 12\times 1$ is used in the Sn-$\sqrt{3}$ BZ and scaled with the size of supercells in magnetic and doped calculations. A Gaussian smearing of $0.001$ Ry is used for all calculations. Ultra soft pseudo-potentials generated using $4$d $5$s $5$p valence electrons for Sn, $3$s $3$p $3$d for Si, $1$s $2$s $2$p for B and $1$s $2$s $2$p for H are used. 
Hybrid HSE06 functional \cite{HSE_2003,HSE_2006} calculations were performed using FHI-AIMS simulation package \cite{FHI-AIMS_2009,FHI-AIMS_2012} with a screening of $0.11$ Bohr$^{-1}$.  We neglected spin-orbit coupling as it is not relevant for in this system \cite{Badrtdinov_2016,Zahedifar_2019}.\\
In TB calculations, evaluations of the spectral function $A\left( \boldsymbol{k},\omega \right)$ are done using a Gaussian energy smearing of $0.03$ eV around quasi-particle energies while DOS calculations were done with a $10\times 10$ k-space grid with a Gaussian smearing of $0.03$ eV.

\subsection{The model for CPT}

We model the system by a minimal TB
model on the triangular lattice, in the presence of a local Hubbard repulsion, as described by the
following Hamiltonian:
\bea\label{lattice_model}
H&=&H_{tb}+H_U
\eea
\bea
H_{tb}&=&\sum_{\vec{r}\vec{r}'}t\left(|\vec{r}-\vec{r}'|\right)
c^{\dagger}_{\vec{r}}c_{\vec{r}'},\\
H_U&=&U_H\sum_{\vec{r}}\hat{n}_{\vec{r}\uparrow}\hat{n}_{\vec{r}\downarrow},
\eea
where: $\vec{r}$ are the lattice sites,
$c_{\vec{r}}=\left(c_{\vec{r}\uparrow},c_{\vec{r}\downarrow}\right)^T$, $c_{\vec{r}\sigma}$ is the operator for the annihilation of one electron on the site $\vec{r}$ with spin $\sigma$,
$t$ are the hopping amplitude, $\vec{\sigma}=(\sigma_x,\sigma_y,\sigma_z)$
are the Pauli's matrices and $\hat{n}_{\vec{r}\sigma}=c^\dagger_{\vec{r}\sigma}c_{\vec{r}\sigma}$.
Note that the amplitudes of the non-local one-particle interactions,
$t$ depend only on the distance. In what follows we consider interactions up to the $6^{\text{th}}$ nearest neighbor.
The numerical values of the amplitudes
$t$ are obtained by fitting the non-interacting (Non-magnetic) DFT bands in \ref{DFT_undoped}.

\begin{figure}
\centering
\includegraphics[width=\columnwidth]{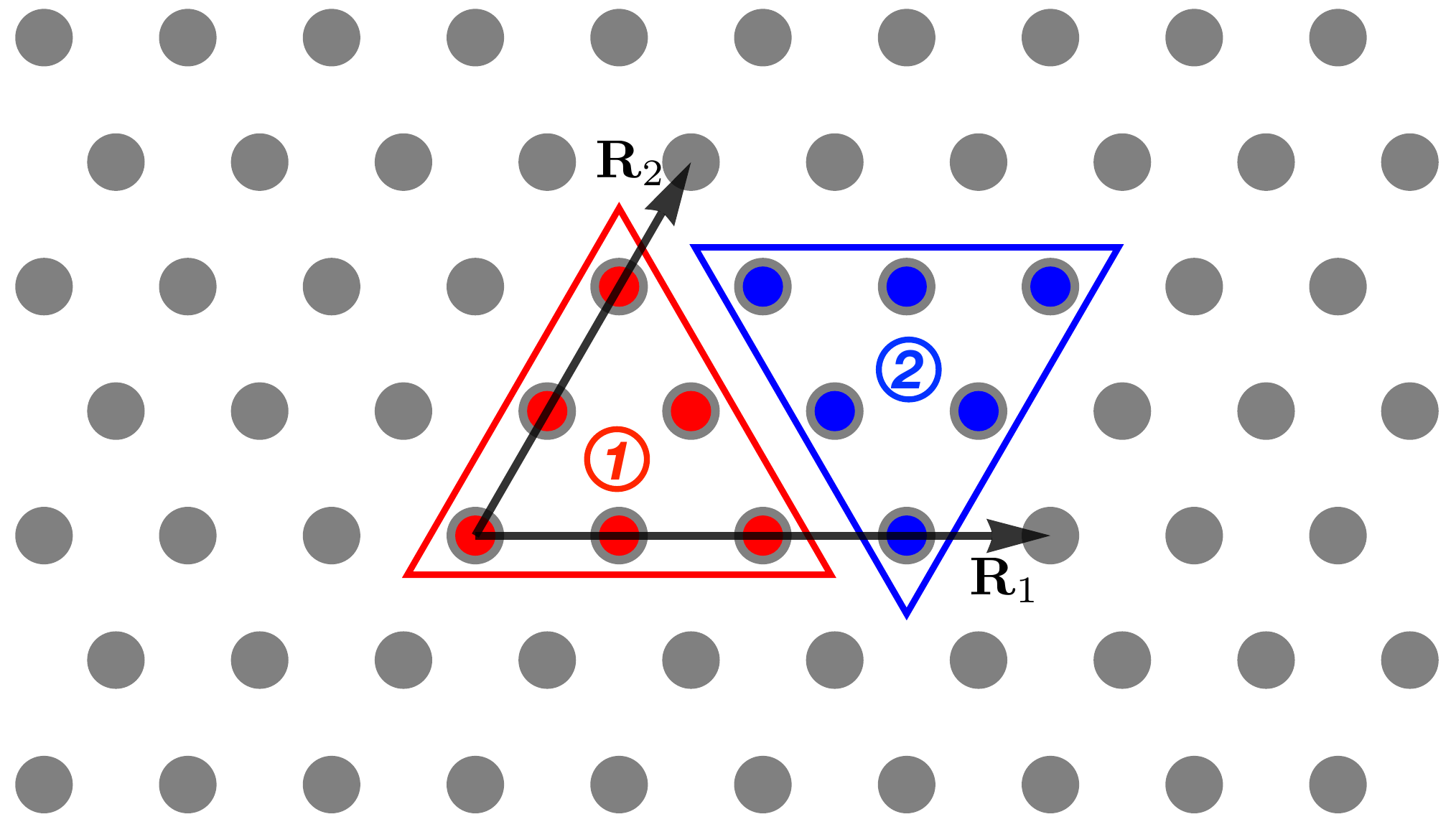}
\caption{
Sketch of the cluster arrangement used for the implementation of the CPT.
The exact diagonalization is performed in each of the two clusters, 1(red) and 2(blue), by using open boundary conditions. The supercell defined by the CPT is given by the ensemble of cluster 1 and 2, so that the $\mathbf{R}_1$ and $\mathbf{R}_2$ are the unit vectors of the superlattice. 
}
\label{fig_clusters}
\end{figure}

Once the lattice model, Eq. \pref{lattice_model},
has been defined, we proceed by applying the CPT,
as described briefly below
(for a review of the CPT's, see the Ref. \cite{Senechal_prb2002}).
We tile the full lattice by joining the two 6-atoms triangular clusters shown in the Fig.\ref{fig_clusters}.
Note that using two paired inverted clusters is
necessary for the tiling.
The supercell defined by the CPT is given by the ensemble of cluster 1 and 2, so that $\mathbf{R}_1$ and $\mathbf{R}_2$ in Fig.\ref{fig_clusters} are the unit vectors of the superlattice. 
A similar cluster arrangement has been adopted in the Ref. \cite{Sahebsara_prl2008}.
First, we perform the exact diagonalization of the Hamiltonian \pref{lattice_model} in the $2^{12}$-dimensional Hilbert space of
each single cluster, using open boundary condition.
Next, we compute the frequency dependent local Green's function of the cluster at zero temperature, as given by the Kallen-Lehmann decomposition. Let $\ket{GS},E_{GS},\ket{n},E_n$ be the GS, its energy, the $n$-th excited state and its energy respectively, then the cluster's Green's function is:
\bea\label{cluster_green}
G^{cl}_{\vec{r}\sigma,\vec{r}'\sigma'}(\omega)&=&\sum_n\left\{
\frac{\bra{GS}c_{\vec{r}\sigma}\ket{n}
\bra{n}c^\dagger_{\vec{r}'\sigma'}\ket{GS}}
{\hbar\omega-E_n+E_{GS}}+\right.\\
&+&\left.\frac{\bra{GS}c^\dagger_{\vec{r}'\sigma'}\ket{n}
\bra{n}c_{\vec{r}\sigma}\ket{GS}}
{\hbar\omega+E_n-E_{GS}}
\right\},\nn
\eea
which is a $12\times12$ matrix.
Then we define the inter-cluster Hamiltonian, $T_{12}$,
describing the coupling between the cluster 1 and 2.
Note that $T_{12}$ depends only on the hopping amplitudes, the $t$'s, as the electronic interaction is purely local.
At the first order in $t/U_H$, the local Green's function of the supercell is given by the $24\times24$ matrix:
\bea
G^{loc}(\omega)=\begin{pmatrix}
\left[G^{(1)}(\omega)\right]^{-1}&&-T_{12}\\
-T^\dagger_{12}&&\left[G^{(2)}(\omega)\right]^{-1}
\end{pmatrix}^{-1},
\eea
where $G^{1}$ and $G^{2}$ are the Green's functions corresponding to the cluster 1 and 2, respectively, as defined by the Eq. \pref{cluster_green}.
$G^{loc}$ describes the local dynamics of the single particle in the supercell formed by the clusters 1 and 2. In order to analyze the dynamics in the full crystal, we introduce the coupling Hamiltonian, $\hat{H}_{\mathbf{R},\mathbf{R}'}$, between the supercell at $\mathbf{R}$ and that at $\mathbf{R}'$, where $\mathbf{R},\mathbf{R}'=n_1\mathbf{R}_1+n_2\mathbf{R}_2$, $n_1,n_2$ integers. As it is for $T_{12}$, $\hat{T}_{\mathbf{R},\mathbf{R}'}$ does not depend on the local interaction. In the reciprocal space, the Fourier transform function of $\hat{T}$ is given by:
\bea
\hat{T}(\vec{k})=\sum_{\mathbf{R}}\hat{T}_{\mathbf{R},\mathbf{R}'}e^{-i\vec{k}\cdot\left(\mathbf{R}-\mathbf{R}'\right)},
\eea
where the wave vector $\vec{k}$ is defined in the BZ of the superlattice,
which is $M=12$ times smaller than that of the original lattice (or $M/3$ times smaller in the case of the $3\times3$ reconstruction). At the first order in $t/U_H$, the Green's function as a function of $\omega$ and $\vec{k}$ is given by:
\bea
G^{\text{CPT}}(\vec{k},\omega)=\left\{\left[G^{loc}(\omega)\right]^{-1}-
\hat{T}(\vec{k})\right\}^{-1},
\eea
\begin{figure}
\centering
\includegraphics[width=0.95\columnwidth]{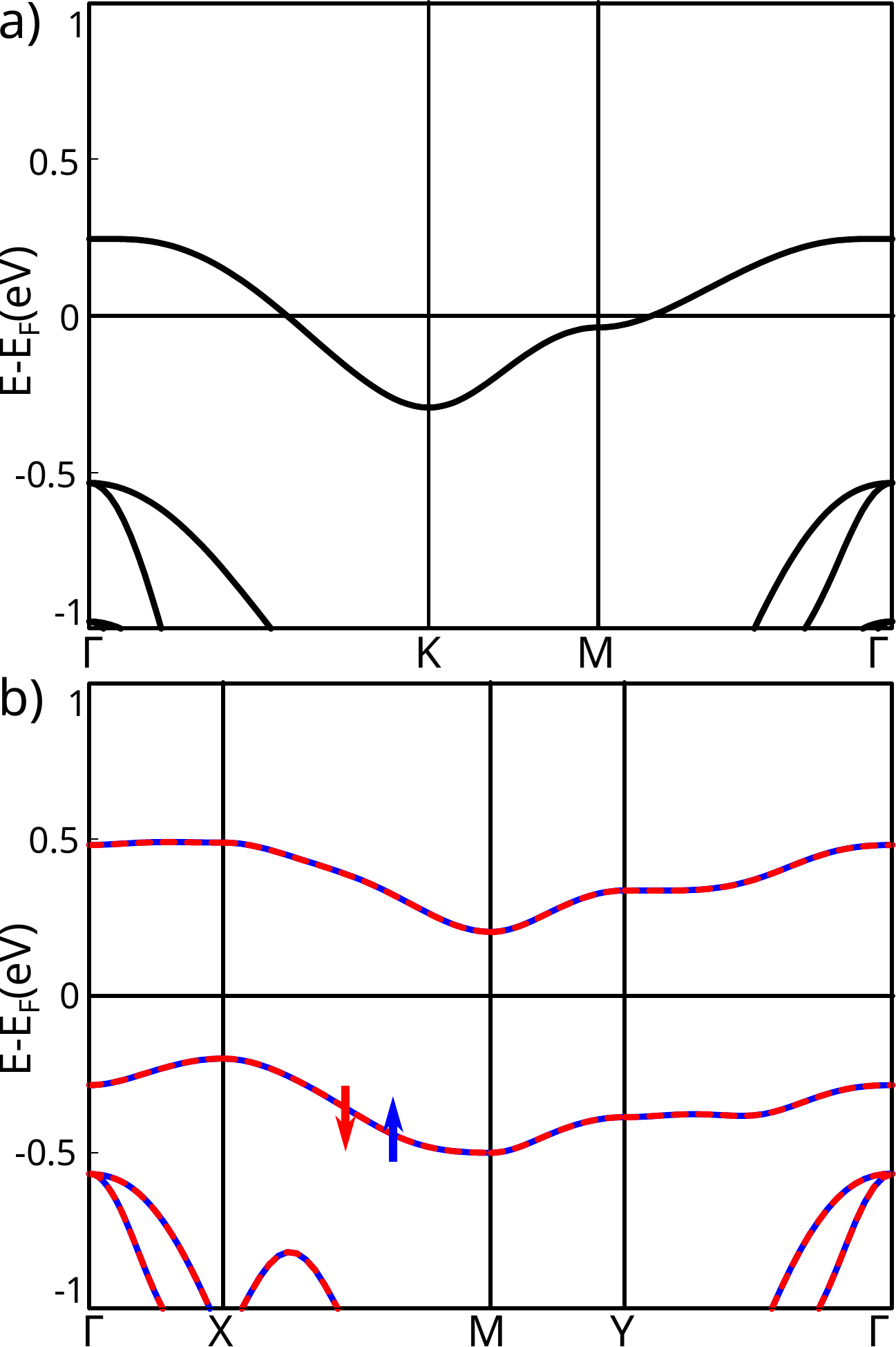}
\caption{
a) \& b) HSE06 calculated band structure in the non-magnetic and anitferromagnetic phases respectively. Red and blue lines in b) correspond to spin down (red arrow) and spin up (blue arrow) bands respectively.
}
\label{DFT_undoped_APPENDIX}
\end{figure}

which is a $24\times24$ matrix with entries: $G^{\text{CPT}}_{\vec{r}\sigma,\vec{r}'\sigma'}(\vec{k},\omega)$,
where $\vec{r},\vec{r}'$ are the internal coordinates of the supercell. Finally, one has to project $G^{\text{CPT}}(\vec{k},\omega)$ in the original BZ, that can be done according to the following prescription\cite{Senechal_prb2002}:
\bea\label{greens_projection}
g^{\text{CPT}}_{\sigma\sigma'}(\vec{k},\omega)=\frac{1}{M}\sum_{\vec{r}\vec{r}'}
G^{\text{CPT}}_{\vec{r}\sigma,\vec{r}'\sigma'}(\vec{k},\omega)e^{-i\vec{k}\cdot\left(\vec{r}-\vec{r}'\right)},
\eea
where now $\vec{k}$ is defined in the original BZ.
Note that the Eq. \pref{greens_projection} is neglecting the Umklapp processes, that describe the transfer of wave vectors belonging to the reciprocal superlattice and are a consequence of the lack of the traslational invariance of $G^{\text{CPT}}_{\vec{r}\sigma,\vec{r}'\sigma'}$ in the supercell.
The single particle excitations are identified by the resonances of the energy-loss function:
\bea\label{energy_loss}
\rho(\vec{k},\omega)=-\frac{\hbar}{\pi}\mathrm{Im}\left\{\mathrm{Tr}\left[g^{\text{CPT}}(\vec{k},\omega+i\gamma)\right]\right\},
\eea
where $\gamma$ is a numerical broadening.
Among all the important informations
of the system that can be extracted from $\rho$, one can compute the Fermi energy, $E_F$, as defined by:
\bea
n=\int_{-\infty}^{E_F/\hbar}\,d\omega\int_{BZ}\frac{\,d^2k}{A_{BZ}}\rho(\vec{k},\omega),
\eea
where $n\in[0,2]$ is the density of electrons and $A_{BZ}=\sqrt{3}a^2/2$ is the area of the BZ. In the numerical applications,
we sample the BZ with a grid of $N_k$ points,
which is also the number of atoms in the lattice, and replace:
\bea
\int_{BZ}\frac{\,d^2k}{A_{BZ}}\rightarrow \frac{1}{N_k}\sum_{\vec{k}}.
\eea

\subsection{DFT hybrid functional calculations}

\begin{figure}
\centering
\includegraphics[width=\columnwidth]{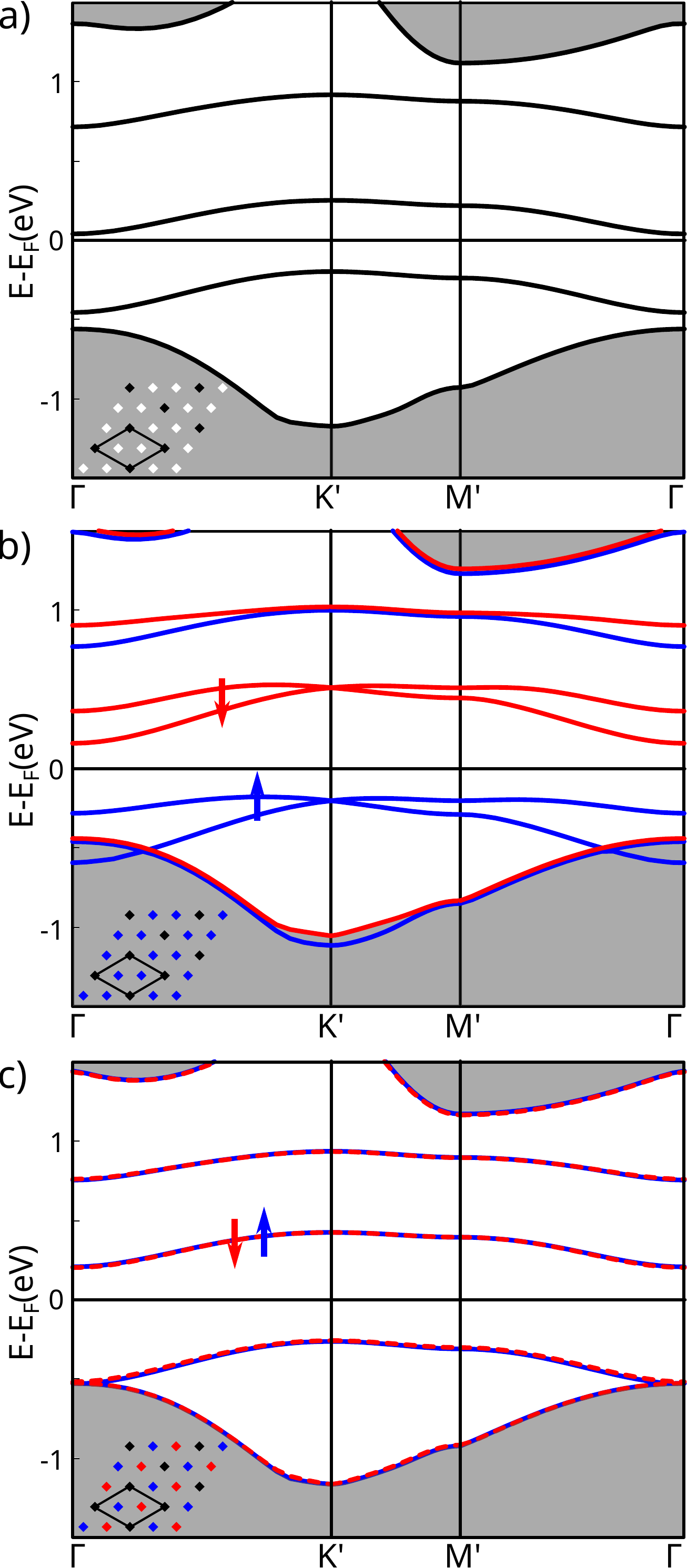}
\caption{
a), b) and c) band structure calculations in the non-magnetic, ferromagnetic and antiferromagnetic supercells respectively done with HSE06 functional. Red and blue lines in b) and c) correspond to spin down (red arrow) and spin up (blue arrow) bands respectively. The insets are sketch of the structure used in the respective band structure calculation: black, white, red and blue circles represent B-doped, non-magnetic, spin up and spin down Sn adatoms.
}
\label{DFT_doped_B_APPENDIX}
\end{figure}

In support of DFT+U calculations we perform an hybrid functional analysis of the system under investigation by means of HSE06 functional. Calculations of the undoped Sn-$\sqrt{3}$ are reported in Fig.\ref{DFT_undoped_APPENDIX}. Qualitatively, the Sn surface bands are almost unchanged with respect to DFT+U. HSE06 calculations gives a slightly higher band width and an higher MI gap ($\sim 0.35$ eV). The Si gap ($\sim 1.77$ eV) is highly overestimated due to the number of silicon bi-layer used.\\
As already commented in the main text, the electronic capture by means of the B atom does not depend on the functional used as it is clear from Fig.\ref{DFT_doped_B_APPENDIX}, where we report the band structure of the doped $3\times 3$ supercell. In HSE06 calculations, the displacement of the Sn adatom over the B atom is the same of DFT+U calculations ($\sim 0.3$ \AA). The energy gaps are generally higher in HSE06 with respect to the one computed in DFT+U. NM calculations in the doped $3\times 3$ (see Fig.\ref{DFT_doped_B_APPENDIX}a) gives an insulating GS caused by a warping of a second Sn adatom towards the surface as described in \ref{APPENDIX_charge_order}.

\subsection{ A boron-induced non-magnetic charge order}\label{APPENDIX_charge_order}

\begin{figure}
\centering
\includegraphics[width=\columnwidth]{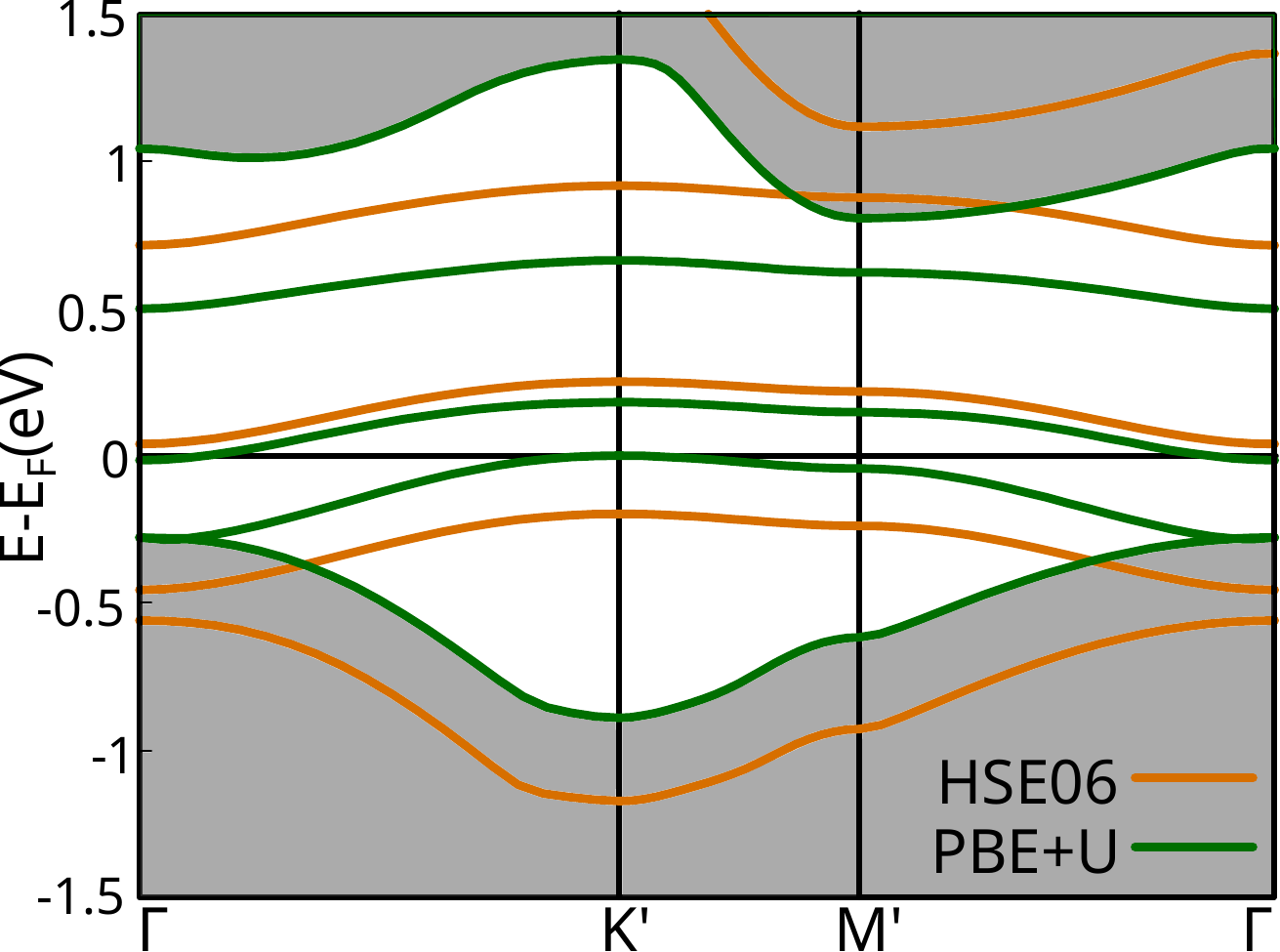}
\caption{
HSE06 (orange) and DFT+U (green) non-magnetic band structure calculations of the $3 \times 3$ supercell with one boron atom ($\theta_B=0.11$ML). 
}
\label{DFT_NM_CDW}
\end{figure}

During B-doped DFT calculations we noticed that NM calculations done in the $3\times 3$ supercell ($\sim 0.1$ ML B coverage), give an unexpected insulating GS. The insulating behavior, is given by an adjustment of the height of a second Sn adatom towards the silicon surface. 
Similarly to the Sn-B charge capture, this distortion comes with a partially electronic transfer from the warped to the upper Sn adatom that causes the system to be insulating (see Fig.\ref{DFT_NM_CDW}). 
Similar charge orders have been found in analogue pristine compounds such as Pb/Ge(111) \cite{Goldoni_prb97,Ren_prb16,Mascaraque_prb98}, Sn/Ge(111) \cite{Cortes_2013} and Pb/Si(11) \cite{Ren_prb16,PhysRevB.107.035125,Tresca_2018} but not in the pristine Sn/Si(111), both experimentally and theoretically. The simulated doping allows Sn/Si(111) to undergo an insulating charge ordering that is peculiar to the simulated $3\times 3$ supercell. We wish to highlight that in HSE06 calculations this charge ordering is found as a unique outcome of the ionic relaxation, even if the starting geometry is symmetric. In DFT+U calculation the relaxation has to be "guided" by starting with an asymmetric structure in order to find the lower-in-energy (of about $15$ meV with respect to the symmetric structure) charge order, underlining a local minimum in the energy landscape as a function of the Sn z-displacing of the PBE+U functional.\\

\section{Experimental details}
\label{APPENDIX_EXP}
\begin{figure}
\includegraphics[width=\columnwidth]{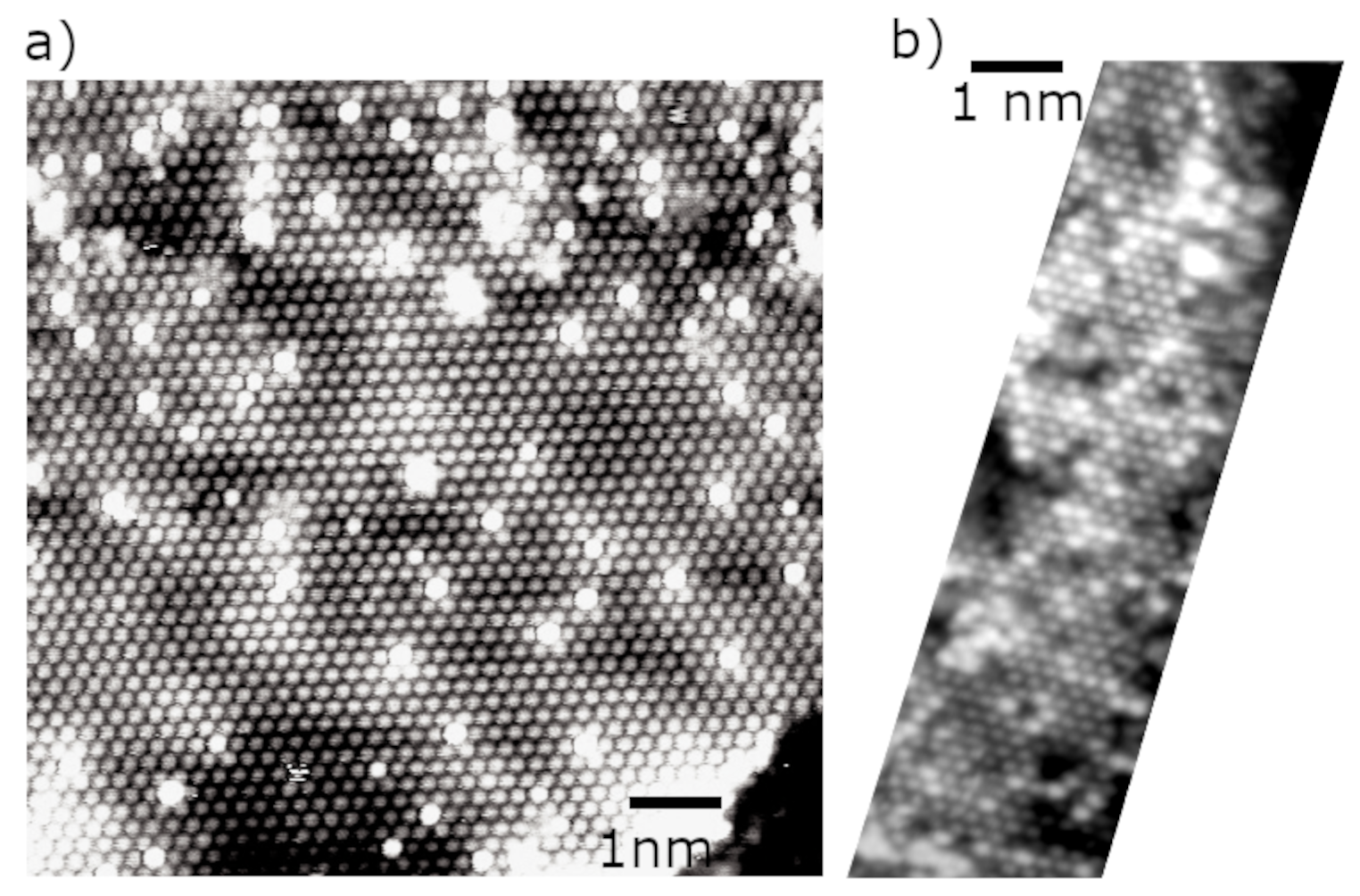}
\caption{STM topographic images of the B-doped Si/Si(111) surfaces used as substrates of the Sn overlayer before the Sn deposition. a) Near B saturation only a few Si adatoms do not have B atoms below immediately them (bright adatoms) and the apparent corrugation caused by deeper B dopants produces darker and brighter areas in the Si overlayer. b) At itermediate B coverages the bright Si adatoms tend to form short disordered chains similarly to Sn adatoms in Fig.\ref{fig1new} d, indicating that the B sites are not affected by the replacement of Si with Sn. 
}
\label{Fig_STMapp}
\end{figure}

The structure of the B-doped Si(111) surfaces before the deposition of the Sn overlayer is shown in the STM topographic images of Fig.\ref{Fig_STMapp} for medium and high B coverage. At medium coverage the Si adatoms without B below (bright adatoms) tend to form short chains \cite{lyo_1989} similarly to the Sn atoms in the Sn terminated surface (see Fig.\ref{fig1new}d)). This is consistent with the expectation that the sites occupied by the B atoms do not change appreciably when Sn replace Si in the topmost layer, and that in both cases the bright adatoms mark the positions of the missing subsurface B atoms. Near saturation B coverage only a few Si adatoms without B are left (isolated bright adatoms) and a smooth corrugation of the order of 0.04 nm is observed in the lattice of the other adatoms at negative sample bias. This corrugation is the result of the electrostatic potential of the B dopant deeper than those in the first bilayer (from 0.6 to about 1.5 nm below the surface) \cite{Makoudi_2017}
and indicates that a few percent of the surface unit cells have a B atom that is not in the S\textsubscript{5} site. The dopant density in first nm below the first bilayer in our samples is estimated to be of the order of 10\textsuperscript{20} cm\textsuperscript{-3} in agreement with References \cite{Makoudi_2017, shen_94}.

\begin{figure}
\includegraphics[width=0.8\columnwidth]{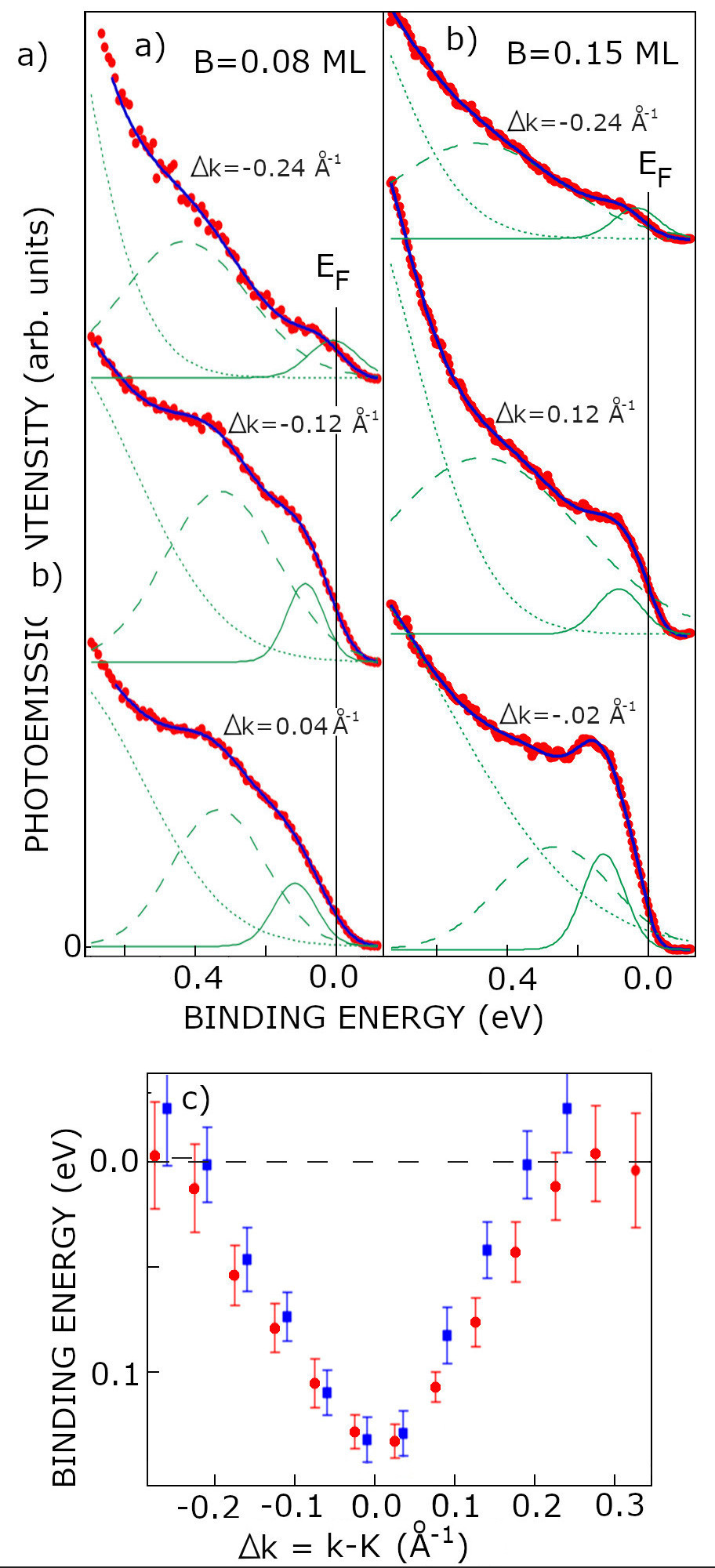}
\caption{Fit of the Fermi level region of the photoemission spectra at low a) and medium b) B coverages for different values $\Delta$ K of the difference between the wavevector and the K point. We used the three Gaussian peaks shown in the figures for the backbond states (dotted line), lower Hubbard band (dashed line) and quasi-particle states (solid line). c) Energy of the quasi-particle band as a function of the wavevector measured from the K point as obtained from the fitting procedure at $\approx 0.08$ ML ( squares) and 0.15 ML (circles) B coverage.
}
\label{bande_fit2.jpg}
\end{figure}

In order to extract quantitative information on the dispersion of the quasi-particle band of the Sn-$\sqrt{3}$ surface the photoemission spectra have been fitted near the Fermi level with the sum of three Gaussians multiplied by the Fermi distribution function and convoluted with a fourth Gaussian representing the energy broadening caused by the surface photovoltage effect and the energy analyzer. The first three Gaussians stand for the Sn backbonds states, the lower Hubbard band and the quasiparticle states. Examples of the results of the fitting are shown in Fig.\ref{bande_fit2.jpg}. The peak between -0.3 and -0.4 eV, the lower Hubbard band, is evident in the spectra at low B coverage, and is needed to fit the spectra also at high coverage, where its presence is less noticeable in the row photoemission data. The quasi-particle peak within 0.2 eV from $E_F$ has a full width at half-maximum of about 0.09 eV and is below the Fermi level only for wavevectors that differ by less than 0.22 \AA \textsuperscript{-1} from K. The dispersion obtained by the fit is shown in Fig.\ref{bande_fit2.jpg} c. The addition of a smooth background to the three peaks does not change the dispersion.

\begin{figure}
\includegraphics[width=\columnwidth]{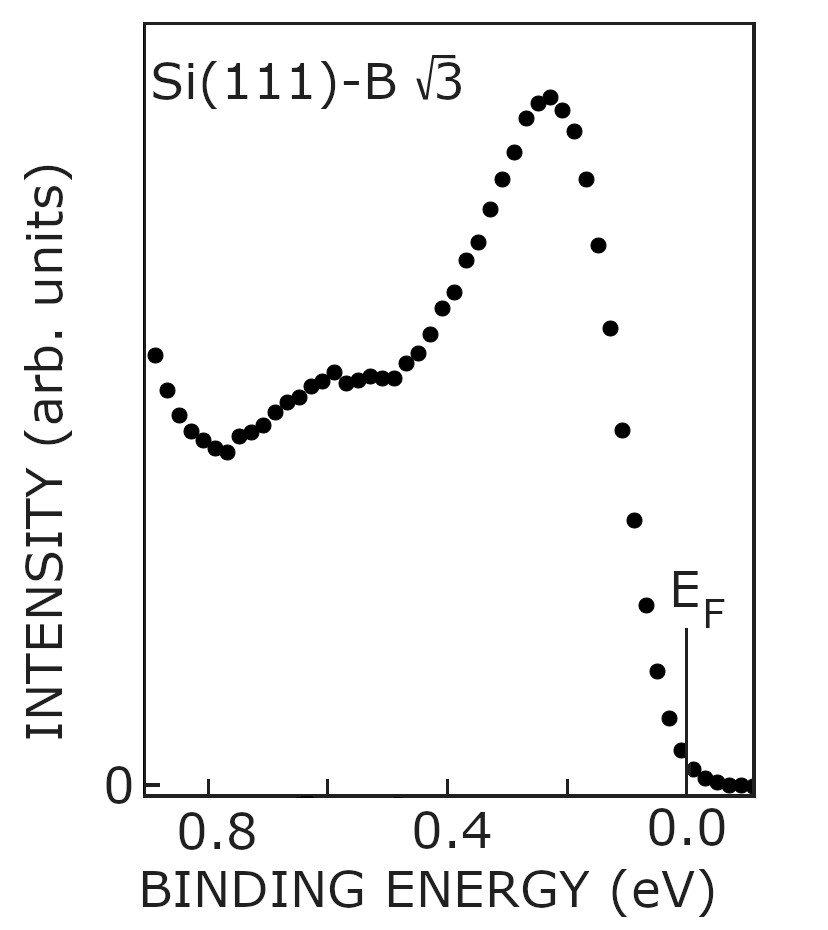}
\caption{Photoemission spectrum of the doped Si-terminated $\sqrt{3}$ surface at 0.2 \AA$^{-1}$ from K at 30 K and for 0.15 B coverage. Only the lower Hubbard band is visible near the Fermi level.
}
\label{Fig_19_B}
\end{figure}

The photoemission spectrum of the B doped Si-$\sqrt{3}$ surface never shows the peak at the Fermi level of the Sn terminated surface, as shown in Fig.\ref{Fig_19_B}. The peak of the lower Hubbard band is 0.25 eV below $E_F$ and the residual intensity at $E_F$ is caused by the high-energy tail of the peak broadened by dishomogeneity of surface photovoltage effect.

\begin{figure}
\centering
\includegraphics[width=0.8\columnwidth]{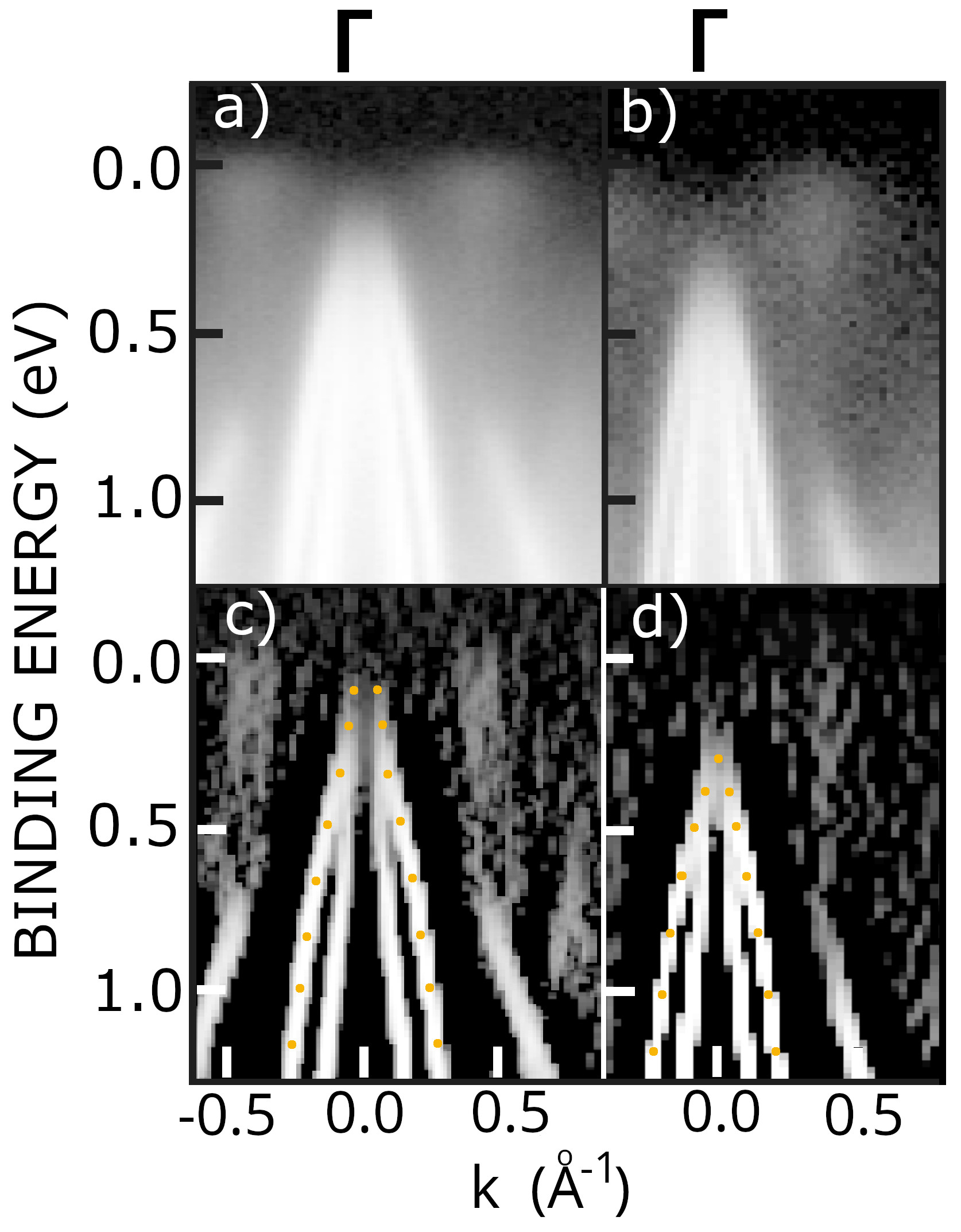}
\caption{Photoemission intensity map of the top of the Si valence band in the third BZ along $\Gamma$M from the B-doped Sn/Si(111)-$\sqrt{3}$ (a) and from the same surface before the Sn layer deposition (b) in a logaritmic scale. Panels c and d are the opposite of the second derivative with respect to k of same data to reduce the background. The scale is logaritmic. The growth of the Sn layer causes the valence band to rise by about 0.2 eV and to reach the Fermi level.
}
\label{banaval.jpg}
\end{figure}

\begin{figure}
\centering
\includegraphics[width=0.8\columnwidth]{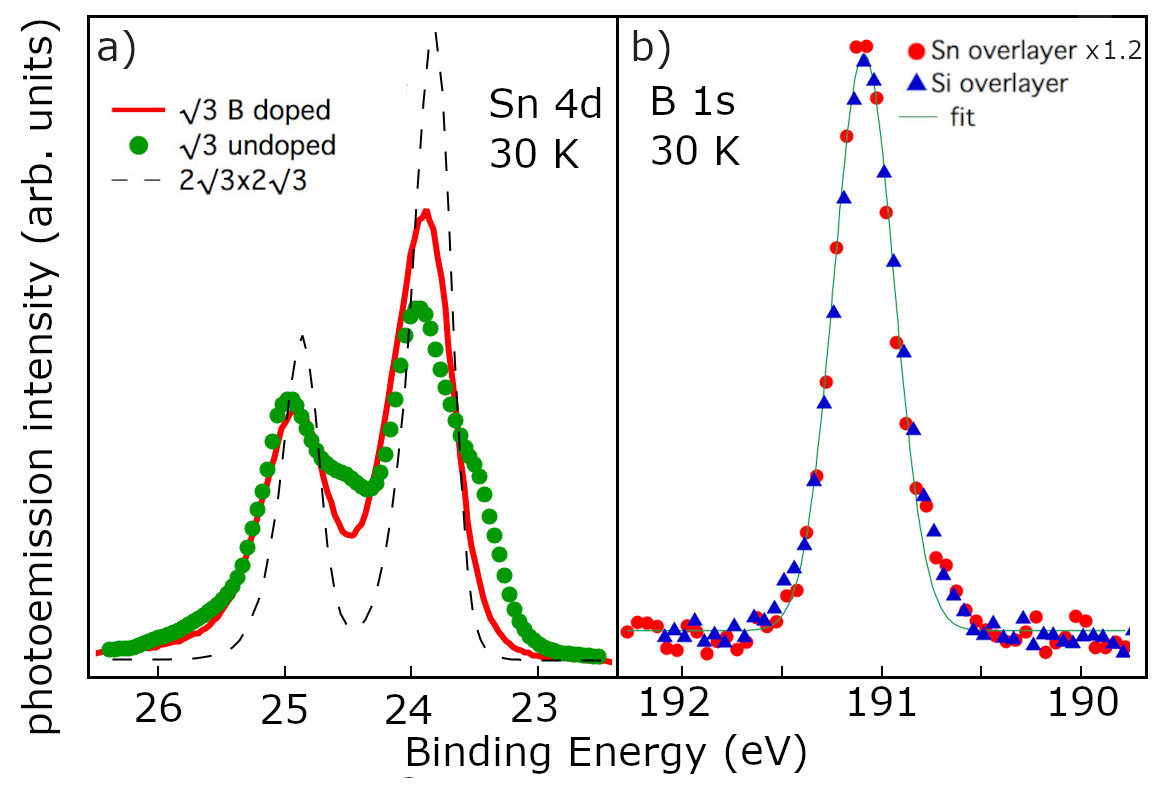}
\caption{a) Sn 4d core levels photoemission of the doped Sn overlayer compared to those of the undoped Sn-$\sqrt{3}$ surface and of the Sn $2\sqrt{3}\times2\sqrt{3}$ overlayer. The energy splitting of the core levels is reduced by the doping, but non removed. The peaks have been nomalized to the same area. The photon energy is 135 eV. b) B 1s core levels of the Si-$\sqrt{3}$ and Sn-$\sqrt{3}$ surfaces normalized to the same intensity. The peaks are  similar and that of the Sn-terminated surface is about 20\% weaker. A Gaussian fit is also shown. The photon energy is 220 eV. }
\label{coreSn2.png}
\end{figure}

A peculiar characteristic of the doped Sn-$\sqrt{3}$ surface is the partial depletion of the Si valence band, as shown by Fig.\ref{banaval.jpg}. This figure compares the top of the valence band of the Si-$\sqrt{3}$ surface (panels b and d) with that of the same surface after Sn deposition (panels a and c). While in the first case the top of the valence band is well below $E_F$, the substitution of Si with Sn  cause the top to disappear above $E_F$. A valence band so close to $E_F$ is not observed on other tetravalent adatom lattices on Si(111) or Ge(111).
The B 1s core level of the B-doped Si-$\sqrt{3}$ surface at high coverage shows a single symmetric and narrow peak \cite{Ma}, as also reported in Fig.\ref{coreSn2.png}, in agreement with the picture of a single site (S\textsubscript{5}) predominantly occupied. The same peak on the Sn-$\sqrt{3}$ surface has the same binding energy and the same width, supporting the idea that the Sn deposition on the B-doped substrate and the subsequent annealing at 520°C do not displace the the vast majority of B atom from the position they had in the Si surface. The main effect of the formation of the Sn $\sqrt{3}$ overlayer on the Si(111)-B $\sqrt{3}$ surface is an approximately 20\% attenuation of the intensity of the B core peak. This is consistent with the picture of the formation of regions where an additional Si bilayer is inserted between the B layer and the Sn layer (see section II A). A full Si bilayer would attenuate the signal approximately by a factor 2 at the low kinetic energies of the B1s electrons in our spectra, i.e. about 30 eV.

The 20\% attenuation would correspond to about 40\% of the surface with the extra bilayer. Since this bilayer is formed by 1/3 ML of Si adatoms displaced by the Sn atoms it should cover 17\% of the surface. Differences in the photoelectron diffracion effects caused by the substitution of Si with Sn may be one reason of this discrepancy.
The Sn 4d core levels of the undoped Sn-$\sqrt{3}$ surface show two or more doublets interpreted as result of charge fluctuations of a strongly correlated state \cite{Hansmann_2016} (see Fig.\ref{coreSn2.png}).
The same core levels of the B doped surface are appreciably narrower, even if the doping introduces chemical inhomogeneities with nonequivalent Sn sites in the Sn-$\sqrt{3}$ phases and the amorphous phase. This latter phase has a small coverage at low B doping, therefore it should not be responsible for the narrowing observed in Fig.\ref{coreSn2.png}.
There is a residual broadening of the Sn 4d levels of the doped Sn-$\sqrt{3}$ phase since those of the  $2\sqrt{3}\times2\sqrt{3}$ phase are narrower, even if the latter  contains several kind of nonequivalent Sn sites. Therefore, the Sn 4d spectrum suggests that the charge fluctuations are appreciably modified in the doped phase, but not suppressed. 

\clearpage

\bibliography{Literature}

\end{document}